%% file: 0-main.tex
\RequirePackage{fix-cm}
\documentclass[twocolumn]{svjour3}          
\makeatletter
\if@twocolumn
  \renewcommand\normalsize{%
   \@setfontsize\normalsize\@xpt{12.5pt}%
   \abovedisplayskip=3 mm plus6pt minus 4pt
   \belowdisplayskip=3 mm plus6pt minus 4pt
   \abovedisplayshortskip=0.0 mm plus6pt
   \belowdisplayshortskip=2 mm plus4pt minus 4pt
   \let\@listi\@listI}%

  \renewcommand\small{%
   \@setfontsize\small{8.5pt}\@xpt
   \abovedisplayskip 8.5\p@ \@plus3\p@ \@minus4\p@
   \abovedisplayshortskip \z@ \@plus2\p@
   \belowdisplayshortskip 4\p@ \@plus2\p@ \@minus2\p@
   \def\@listi{\leftmargin\leftmargini
               \parsep 0\p@ \@plus1\p@ \@minus\p@
               \topsep 4\p@ \@plus2\p@ \@minus4\p@
               \itemsep0\p@}%
   \belowdisplayskip \abovedisplayskip}

\else
  \if@smallext
   \renewcommand\normalsize{%
   \@setfontsize\normalsize\@xpt\@xiipt
   \abovedisplayskip=3 mm plus6pt minus 4pt
   \belowdisplayskip=3 mm plus6pt minus 4pt
   \abovedisplayshortskip=0.0 mm plus6pt
   \belowdisplayshortskip=2 mm plus4pt minus 4pt
   \let\@listi\@listI}%

  \renewcommand\small{%
   \@setfontsize\small\@viiipt{9.5pt}%
   \abovedisplayskip 8.5\p@ \@plus3\p@ \@minus4\p@
   \abovedisplayshortskip \z@ \@plus2\p@
   \belowdisplayshortskip 4\p@ \@plus2\p@ \@minus2\p@
   \def\@listi{\leftmargin\leftmargini
               \parsep 0\p@ \@plus1\p@ \@minus\p@
               \topsep 4\p@ \@plus2\p@ \@minus4\p@
               \itemsep0\p@}%
   \belowdisplayskip \abovedisplayskip}
 \else
  \renewcommand\normalsize{%
   \@setfontsize\normalsize{9.5pt}{11.5pt}%
   \abovedisplayskip=3 mm plus6pt minus 4pt
   \belowdisplayskip=3 mm plus6pt minus 4pt
   \abovedisplayshortskip=0.0 mm plus6pt
   \belowdisplayshortskip=2 mm plus4pt minus 4pt
   \let\@listi\@listI}%

  \renewcommand\small{%
   \@setfontsize\small\@viiipt{9.25pt}%
   \abovedisplayskip 8.5\p@ \@plus3\p@ \@minus4\p@
   \abovedisplayshortskip \z@ \@plus2\p@
   \belowdisplayshortskip 4\p@ \@plus2\p@ \@minus2\p@
   \def\@listi{\leftmargin\leftmargini
               \parsep 0\p@ \@plus1\p@ \@minus\p@
               \topsep 4\p@ \@plus2\p@ \@minus4\p@
               \itemsep0\p@}%
   \belowdisplayskip \abovedisplayskip}
  \fi
\fi
\let\footnotesize\small
\makeatother
\smartqed  
\usepackage{graphicx}
\usepackage{amsmath}
\usepackage{amssymb}

\usepackage{blindtext}
\usepackage{caption, subcaption}
\usepackage{multirow}
\usepackage{booktabs}
\usepackage{verbatim}
\usepackage{color, colortbl}
\usepackage{xspace, xcolor}
\usepackage{cite}
\usepackage[export]{adjustbox}
\usepackage{comment}
\usepackage{soul}
\definecolor{citecolor}{RGB}{34,139,34}

\usepackage{booktabs}
\usepackage{amsmath}
\usepackage{amssymb}
\usepackage{mathtools}
\usepackage{xspace}
\usepackage{hyperref}
\usepackage{graphicx}
\usepackage{caption}
\usepackage{subcaption}
\usepackage{multirow}
\usepackage{color, colortbl}
\usepackage{fancyvrb}
\usepackage{xcolor}
\usepackage{appendix}
\usepackage{natbib}

\definecolor{Gray}{gray}{0.9}

\usepackage{microtype} 
\usepackage{threeparttable}
\hypersetup{pdfstartview={FitH}}
\hypersetup{pdfborder={0 0 0}}
\hypersetup{pdfpagemode={UseNone}}
\hypersetup{colorlinks=true}
\hypersetup{citecolor=blue}
\hypersetup{linkcolor=blue}

\begin{document}

\title{On Measuring and Controlling the Spectral Bias\\of the Deep Image Prior
}


\author{Zenglin~Shi$^1$        \and
        Pascal~Mettes$^1$       \and
        Subhransu~Maji$^2$      \and
        Cees~G.~M.~Snoek$^1$
}

\authorrunning{Zenglin~Shi et al.} 

\institute{
Zenglin~Shi  \\
\email{z.shi@uva.nl} \\ \\
Pascal Mettes  \\
\email{P.S.M.Mettes@uva.nl} \\ \\
Subhransu Maji  \\
\email{smaji@cs.umass.edu} \\ \\
Cees Snoek  \\
\email{cgmsnoek@uva.nl} \\ \\
$^1$ University of Amsterdam \\
$^2$ University of Massachusetts, Amherst
}

\date{Received: date / Accepted: date}

\maketitle
\input{notation}
\input{1-abstract}

\input{2-introduction}
\input{3-related-work}
\input{4-method}
\input{5-experiment}
\input{6-conclusion}


%
%

\bibliographystyle{spbasic}      
\bibliography{egbib}   

\end{document}

%% file: notation.tex
\def\eg{\textit{e.g.}}
\def\ie{\textit{i.e.}}
\def\Eg{\textit{E.g.}}
\def\etal{\textit{et al. }}
\def\etc{\textit{etc.}}
\newcommand{\dimny}{\mathcal{M}\xspace}
\newcommand{\dimyH}{\mathcal{H}\xspace}
\newcommand{\dimyW}{\mathcal{W}\xspace}
\newcommand{\dimH}{\ensuremath{H}}
\newcommand{\dimW}{\ensuremath{W}}
\newcommand{\dimC}{\ensuremath{C}}
\newcommand{\nStage}{\mathcal{L}\xspace}
\newcommand{\scale}{\mathcal{S}\xspace}
\newcommand{\mypartitletwo}[2][2]{\vspace*{-#1 ex}~\\{\noindent {\bf #2}}}
\newcommand{\mypartitle}[1]{\vspace*{-3ex}~\\{\noindent \underline{\bf #1}}}
\newcommand{\todo}[1]{\textcolor{red}{\textbf{#1}}}
\newcommand{\dimn}{\ensuremath{M}}
\newcommand{\apriori}{\textit{a priori}\xspace}
\newcommand{\mapping}{\ensuremath{G}\xspace}
\newcommand{\params}{\ensuremath{\theta}\xspace}
\newcommand{\data}{\ensuremath{X}\xspace}
\newcommand{\SV}{\ensuremath{X}\xspace}
\newcommand{\pro}{\ensuremath{P}\xspace}
\newcommand{\gt}{\ensuremath{G}\xspace}
\newcommand{\npro}{\ensuremath{N}\xspace}
\newcommand{\featSpace}{\ensuremath{\mathrm{\cal X}}\xspace}
\newcommand{\lSpace}{\ensuremath{\mathrm{\cal Y}}\xspace}
\newcommand{\labbb}{\ensuremath{\mathbf{t}}\xspace}
\newcommand{\state}{\ensuremath{z}\xspace}
\newcommand{\nframes}{\ensuremath{T}\xspace}
\newcommand{\kupdate}{\ensuremath{\boldsymbol{\varphi}}\xspace}
\newcommand{\sol}{\ensuremath{\boldsymbol{\beta}}\xspace}
\newcommand{\nsamples}{\ensuremath{N}\xspace}
\newcommand{\Msamp}{\ensuremath{M_{\mathrm{s}}}\xspace}
\newcommand{\nparticles}{\ensuremath{P}\xspace}

\newcommand{\nDepth}{\ensuremath{D_{\mathrm{max}}}\xspace}
\newcommand{\nTrees}{\ensuremath{K}\xspace}
\newcommand{\Xmat}{\ensuremath{\mathbf{X}}\xspace}
\newcommand{\Ymat}{\ensuremath{\mathbf{Y}}\xspace}
\newcommand{\HH}{\ensuremath{\mathbf{H}}\xspace}
\newcommand{\Smat}{\ensuremath{\mathbf{S}}\xspace}
\newcommand{\Dmat}{\ensuremath{\mathbf{D}}\xspace}
\newcommand{\eye}{\ensuremath{\mathbf{e}}\xspace}
\newcommand{\err}{\ensuremath{\boldsymbol{\xi}}\xspace}
\newcommand{\coeff}{\ensuremath{\mathbf{w}}\xspace}
\newcommand{\samp}{\ensuremath{\mathbf{x}}\xspace}
\newcommand{\laby}{\ensuremath{\mathbf{y}}\xspace}
\newcommand{\func}{\ensuremath{\mathbf{g}}\xspace}
\newcommand{\thresh}{\ensuremath{\tau}\xspace}
\newcommand{\treedepth}{\ensuremath{\Gamma_{\mathrm{depth}}}\xspace}
\newcommand{\sampler}{\emph{Sampler}}
\newcommand{\normal}{\ensuremath{\mathrm{\cal N}}\xspace}
\newcommand{\ssvmCost}{\ensuremath{\ell}\xspace}
\newcommand{\Perp}{\perp \! \! \! \perp}
\def\ci{\perp\!\!\!\perp}
\newcommand{\RR}{I\!\!R}

\newcommand{\labl}{\ensuremath{y}\xspace}

\newcommand{\mean}{\ensuremath{\mu}\xspace}

\newcommand{\Mult}{\ensuremath{\mbox{Mult}}\xspace}
\newcommand{\Cat}{\ensuremath{\mbox{Categorical}}\xspace}
\newcommand{\argmin}{\mathop{\mathrm{arg\,min}}}
\newcommand{\argmax}{\mathop{\mathrm{arg\,max}}}

\newcommand{\cs}[1]{\textcolor{red}{[\textbf{CS}: #1]}}
\newcommand{\psmm}[1]{\textcolor{orange}{[\textbf{PM}: #1]}}
\newcommand{\zl}[1]{\textcolor{blue}{[\textbf{ZL}: #1]}}
\newcommand{\sm}[1]{\textcolor{red}{[\textbf{SM}: #1]}} 

%% file: 1-abstract.tex
\begin{abstract}
The deep image prior showed that a randomly initialized network with a suitable architecture can be trained to solve inverse imaging problems by simply optimizing it's parameters to reconstruct a single degraded image.
However, it suffers from two practical limitations.
First, it remains unclear how to control the prior beyond the choice of the network architecture. Second, training requires an oracle stopping criterion as during the optimization the performance degrades after reaching an optimum value. To address these challenges we introduce a frequency-band correspondence measure to characterize the spectral bias of the deep image prior, where low-frequency image signals are learned faster and better than high-frequency counterparts. Based on our observations, we propose techniques to prevent the eventual performance degradation and accelerate convergence. We introduce a Lipschitz-controlled convolution layer and a Gaussian-controlled upsampling layer as plug-in replacements for layers used in the deep architectures. The experiments show that with these changes the performance does not degrade during optimization, relieving us from the need for an oracle stopping criterion. We further outline a stopping criterion to avoid superfluous computation. Finally, we show that our approach obtains favorable results compared to current approaches across various denoising, deblocking, inpainting, super-resolution and detail enhancement tasks. Code is available at \url{https://github.com/shizenglin/Measure-and-Control-Spectral-Bias}.
\keywords{Spectral bias \and Deep image prior}
\end{abstract}

%% file: 2-introduction.tex
\section{Introduction}
This paper considers the problem of inverse imaging, where the task is to recover the original image from the one that is degraded due to noise, blur, down-sampling and other hardships \citep{bertero1998introduction}. This problem is ill-posed, as a degraded image may correspond to several original images. Hence, reconstructing a unique solution that fits the degraded image is difficult, or impossible even, without some prior knowledge about the image or the degradation~\citep{engl1996regularization}.

The classical computer vision approaches to inverse imaging minimize a regularized cost function to incorporate some prior knowledge into the solution, \eg,~\citep{hahn2011orientation,dong2015image,arias2011variational,lin2008limits}. Despite their excellent results, it remains difficult to handcraft an appropriate regularizer and choose a suitable regularisation parameter for a given application because expert knowledge is often required~\citep{ribes2008linear,jin2017deep}. Rather than providing the priors as input, deep neural networks offer the ability to learn image priors from numerous image samples, \eg,~\citep{mccann2017convolutional,lucas2018using, arridge2019solving}. By doing so, the image priors are gradually encoded into network parameters during training and reused in the inference phase. Despite its promise, the dependence on image pairs seen during training may result in poor generalization of the learned priors \cite{zhang2017beyond,zhang2018ffdnet}.

Contrary to the belief that learning on numerous image samples is necessary to obtain useful image priors, \cite{ulyanov2018deep,ulyanov2020deep} show that the architecture of a generator network itself contains an inductive bias independent of learning, where a \emph{deep image prior} can be implicitly captured by a particular network architecture like an encoder-decoder. 
To leverage the deep image prior for solving inverse imaging problems, a suitably designed network is optimized, starting from a random initialization and a random input, on just a single degraded image through gradient descent. The network is able to output a well-restored image, when its optimization is stopped at the right time, with an early-stopping oracle.
The literature studying the deep image prior mostly focuses on designing network architectures \citep{heckel2018deep,cheng2019bayesian,chen2020dip,ho2020neural}.
However, it remains unclear how to control the deep image prior beyond the choice of the network architecture and prevent performance degradation when an oracle to stop the optimization at peak performance is unavailable. In this paper, we study the deep image prior from a complementary perspective to address these problems.

As our first contribution, we study the deep image prior through measuring its spectral bias (Section \ref{sec_measure}). We find that both the networks of the original deep image prior \citep{ulyanov2018deep,ulyanov2020deep} and its variants \citep{heckel2018deep,cheng2019bayesian} exhibit a spectral bias during optimization, where the low frequency components of the target images are learned better and faster than the high-frequency components.
We believe that the spectral bias leads the networks to capture deep image priors during optimization, beyond the choice of the architecture, since natural images are well approximated by low-frequency components according to the power spectrum \citep{simoncelli2001natural}. We measure the spectral bias with a new Frequency-Band Correspondence metric and pinpoint why the performance of the deep image prior gradually degrades after reaching a peak during the optimization.

We observe that deep image prior performance degrades when high-frequency noise is learned beyond a certain level, which could affect the high-frequency image details. As our second contribution, we therefore propose to prevent performance degradation by restricting the ability of the network to fit high-frequency noise (Section \ref{sec_control}). We bound the layers of our network with Lipschitz regularization and introduce a Lipschitz-variant of batch normalization to accelerate and stabilize the optimization.
We also observe that widely used upsampling methods, like bilinear upsampling, over-smooth, which introduces a bias towards lower frequencies. This slows down the learning of the desired higher frequencies, delaying optimization convergence. Therefore, we propose an upsampling method which allows controlling the amount of smoothing and is capable of balancing performance and convergence. Besides these two methods for controlling spectral bias, we further introduce a simple automatic stopping criterion to avoid superfluous computation.

Lastly, we demonstrate the effectiveness of our method on four inverse imaging applications and one image enhancement application: image denoising, JPEG image deblocking, image inpainting, image super-resolution and image detail enhancement (Section \ref{exp_sota}). The experiments show that our method no longer suffers from eventual performance degradation during optimization, relieving us from the need for an oracle criterion to stop early. The automatic stopping criterion avoids superfluous computation. Our method also obtains favorable restoration and enhancement results compared to current approaches, across all tasks.

%% file: 3-related-work.tex
\section{Related Work}

\subsection{Inverse Problems in Imaging}
An inverse problem in imaging is the task of recovering an unknown image $x^* \in \boldsymbol{X}$ from its noisy measurements $y \in \boldsymbol{Y}$, where $y = \mathcal{A}(x^*) + e.$
Here $e \in \boldsymbol{Y}$ denotes some noise in the measurements. The mapping $\mathcal{A}: \boldsymbol{X} \to \boldsymbol{Y}$ denotes the forward operator, which could represent various inverse problems, such as an identity operator for image denoising, convolution operators for image deblurring, 
%
filtered subsampling operators for super-resolution, \etc~Since the operator $\mathcal{A}$ has a non-trivial null space, these inverse problems are often ill-posed. Meaning that the solution is unstable with respect to the measurements, or there are several possible solutions that are consistent with the measurements \citep{bertero1998introduction}. To solve these ill-posed inverse problems, we review the classical knowledge-driven approaches and the recent data-driven approaches with deep neural networks.

The classical knowledge-driven approaches assume some prior knowledge about the image $x^*$, such as smoothness \citep{titterington1985general,katsaggelos1989iterative} or sparsity \citep{daubechies2004iterative,elad2010role}. These approaches typically aim to find a solution that fits well with the measurements $y$ and is consistent with the assumed prior knowledge. To do so, an optimization criterion is used, such as the minimization of the $l_2$ error norm $||y-\mathcal{A}(x^*)||^2$. Then, prior knowledge is incorporated into the solution process through regularization. Specifically, \cite{rudin1992nonlinear} leveraged the fact that in natural images nearby pixels tend to have similar values, and proposed a denoising model with the total variation regularization, which promotes smoothness while preserving edges in images. Based on the finding that natural images can be generally coded by structural primitives such as edges and line segments \citep{olshausen1996emergence}, sparse representation-based regularization models, \eg, \citep{elad2010role,daubechies2004iterative,portilla2009image}, have been successfully used in image deconvolution tasks. A natural image often has many repetitive local patterns, and thus a local image patch always has many similar patches across the image \citep{efros1999texture}. This non-local self-similarity prior was later employed in many inverse imaging problems such as image denoising \citep{dabov2007image}, image deblurring \citep{kindermann2005deblurring} and super-resolution \citep{protter2008generalizing}. Later, \cite{mairal2009non} proposed non-local sparse regularization models which combine the local sparsity and the non-local self-similarity into a unified framework, where the similar image patches are simultaneously coded to improve the robustness of the inverse reconstruction. Despite their excellent results, a downside of these approaches is that their handcrafted regularization only captures a fraction of the prior knowledge about the image, limiting the inverse imaging ability of their models~\citep{ribes2008linear,jin2017deep}.

Data-driven approaches leverage large collections of training data to directly compute regularized reconstructions with deep neural networks.  The central idea is to create a paired dataset of ground truth images $x$ and corresponding measurements $y$, which can be done by simulating (or physically implementing) the forward operator $\mathcal{A}$ on clean data. Subsequently, one can train a network to learn a direct mapping from measurements $y$ to the ground truth images $x$. Most approaches have focused on designing a proper network architecture to learn a high-performing mapping. For example, \cite{dong2015image_pami} learned a convolutional neural network for image super-resolution, and \cite{jain2008natural} learned a convolutional neural network for image denoising. \cite{mao2016image} demonstrated convolution neural networks with encoder-decoder architectures perform better for restoring degraded images. \cite{zhang2017beyond} proposed to use the convolution neural networks with residual blocks and skip connections to further improve image super-resolution and denoising performance.  \cite{ledig2017photo} proposed a generative adversarial network for image super-resolution to recover the finer texture details. \cite{li2018frequency} proposed a computationally efficient frequency domain deep network for image super-resolution. Despite their excellent results, these approaches are sensitive to changes or uncertainty to the forward operator $\mathcal{A}$. For image denoising, for example, a specific network needs to be trained for each considered noise level. To remedy this issue, \cite{lefkimmiatis2018universal} proposed a universal denoising network with non-local filtering layers, which is able to handle a wide range of noise levels using a single set of learned parameters. Recently, \cite{chen2020controllable} proposed a plugin module, which can be inserted into any backbone networks. This plugin allows the once trained network to be used for multiple forward operators in various image processing tasks, including image smoothing, image denoising, image deblocking, image enhancement and neural style transfer. \cite{wan2020bringing} proposed a triplet domain translation network for restoring old photos, in which multiple degradations exist and are mixed. Such supervised approaches typically perform very well but rely on a paired dataset of ground truth images and their measurements, which may not be available. In this work, we consider the unsupervised inverse imaging approach with a deep image prior.

\subsection{Deep Image Prior}
The deep image prior, introduced by \cite{ulyanov2018deep,ulyanov2020deep}, revealed the remarkable ability of untrained convolution neural networks to solve challenging inverse problems by optimizing on just a single degraded image.  Let $f_\theta: \boldsymbol{Z} \to \boldsymbol{Y}$ denote a convolutional neural network parameterized by $\theta \in \Theta$, which transforms a tensor/vector $z \in \boldsymbol{Z}$ to a degraded image $y \in \boldsymbol{Y}$. 
Without training, the network $f_\theta$ has no knowledge about high-level semantic concepts such as the categories of objects in the images. However, the deep image prior found that the network does contain knowledge about the low-level structure of natural images. This prior knowledge is sufficient to model the conditional image distribution $p(x^*|y_0)$. Here, the unknown image $x^*$ has to be determined given a measurement $y_0$, which allows solving inverse problems in imaging. Specifically, we consider energy minimization problems of the type, $\theta^*=  \argmin \limits_\theta E(f_\theta(z);y_0)$ where $E(f_\theta(z);y_0)$ is a task-dependent data term. For inverse imaging problems, $y_0$ is a noisy, low-resolution, compressed, or occluded image. The minimizer $\theta^*$ is obtained using an optimizer such as gradient descent, starting from a random initialization of the parameters. Given a minimizer $\theta^*$ obtained by $N$ steps of gradient descent, we obtain a restoration result by $y^*{=}f_{\theta^*}(z)$. Competitive performance is even feasible when stopping the network optimization with an early-stopping oracle.

The deep image prior has inspired many to investigate how to expand its applications \citep{gandelsman2019double,kattamis2019exploring,rasti2021undip,vu2021deep,dai2020dipdefend}, how to improve its performance \citep{mataev2019deepred,chen2020dip,liu2019image,asim2019patchdip,zukerman2020bp}, how to understand its workings~\citep{ulyanov2018deep, ulyanov2020deep, cheng2019bayesian,heckel2019denoising}, and how to avoid its early-stopping oracle \citep{cheng2019bayesian,heckel2018deep}. 

\cite{liu2019image} and \cite{mataev2019deepred} employ extra regularization to boost performance of the deep image prior. \cite{chen2020dip,ho2020neural} leverage neural architecture search to obtain a better deep image prior network for improved performance. \cite{asim2019patchdip} employ deep image prior on image patches, which improves its reconstruction ability. \cite{zukerman2020bp} improve the deep image prior by using a backprojection loss function. These approaches improve results, but still require an oracle to determine when to stop the optimization. In this paper, we boost the performance of the deep image prior by controlling its spectral bias, and achieve an automatic stopping with a new criterion.

An intuition provided by \cite{ulyanov2018deep,ulyanov2020deep} for the workings of the deep image prior is that their network follows an encoder-decoder architecture, which imposes strong priors about natural images. \cite{heckel2019denoising} further attribute the effects of the deep image prior to the special architecture with convolutions using fixed interpolating filters. Alternatively, \cite{cheng2019bayesian} explain the deep image prior from a Bayesian perspective by showing that the model behaves like a stationary Gaussian process at initialization. These works have focused on studying the workings of deep image prior, mostly from the view of the network architecture design. In this paper, we provide a complementary perspective. We show that the spectral bias leads the networks to capture deep image priors during optimization, beyond the choice of the architecture. We do so by introducing a metric, the Frequency Band Correspondence, which offers a spectral measurement of the deep image prior, revealing the low-frequency natural image signals are learned faster and better than high-frequency noise signals.

A downside of the original deep image prior \citep{ulyanov2018deep,ulyanov2020deep} is the requirement of an oracle to determine when to stop the optimization as its performance degrades after reaching a peak over the iterations of optimization. \cite{heckel2018deep} tackle this problem with an underparameterized network, at the expense of reduced performance. \cite{cheng2019bayesian} avoid the need for early stopping with a Bayesian approach, at the expense of slower convergence. In this paper, we prevent the performance degradation over iterations with Lipschitz-controlled spectral bias and enable stopping the optimization automatically at an appropriate moment with a new criterion. 

A few recent works \citep{rahaman2019spectral,xu2019frequency,chakrabarty2019spectral} have paid attention to the spectral bias as well. \cite{rahaman2019spectral} and \cite{xu2019frequency} analyze the spectral bias for classification problems with supervised learning, not for generative models with a single image. \cite{chakrabarty2019spectral} exposed the deep image prior has a spectral bias by adding noise at different frequencies to the image and analyzing the optimization trajectories from different noisy versions of the input. However, they do not measure and control the bias. In this work, we propose a frequency band correspondence to measure the spectral bias of the deep image prior. We further control the bias to address the performance degradation problem and the performance-convergence trade-off problem.

%% file: 4-method.tex
\begin{figure}[!h]
    \centering
    \includegraphics[width=1\linewidth]{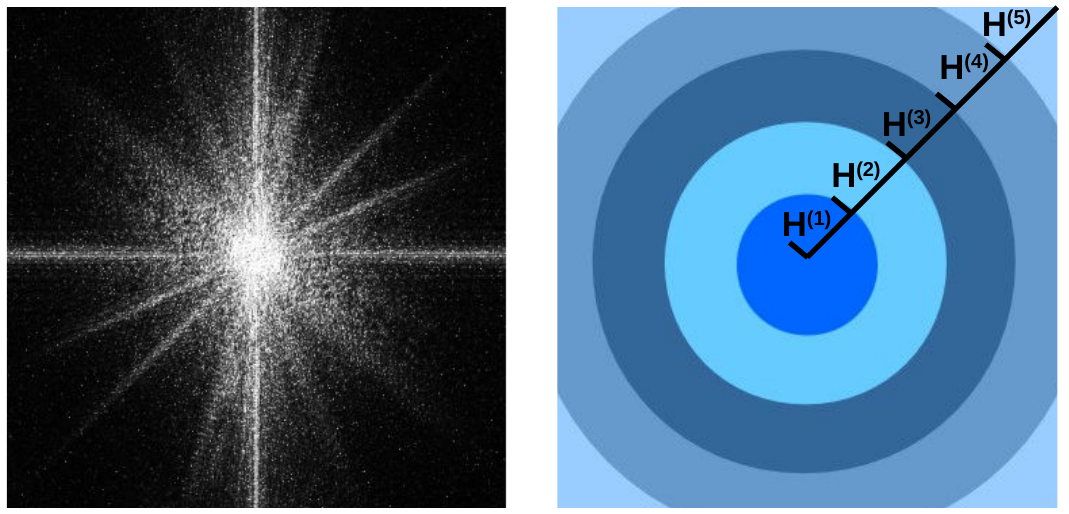}
    \caption{\textbf{Frequency-band correspondence metric.} The left image shows an example of correspondence map $H$, which is computed according to Eq. (\ref{eq:fbc}). We divide the correspondence map into $N$ subgroups corresponding to $N$ non-overlapping frequency bands. Since the correspondence map is symmetrical around the center, we group it according to the distance between its elements and its center uniformly, as illustrated by the right image when $N=5$. Different colors represent different subgroups. We compute the mean correspondence for each band to transform the 2D map to the 1D one.}
    \label{exp_fig_architecture}
\label{fig:fbc}
\end{figure}
\begin{figure*}
    \centering
    \includegraphics{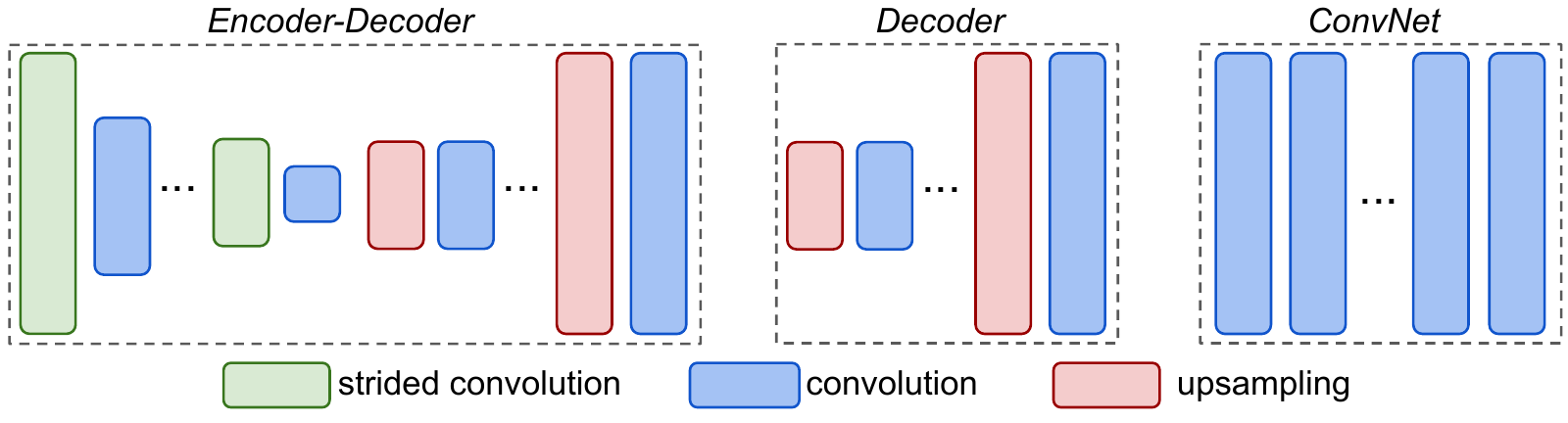}
    \caption{\textbf{Network architectures} used in the experiments of Section \ref{sec_measure}. The \emph{Encoder-Decoder} is the same as the one used in \cite{ulyanov2020deep}. Specifically, the encoder contains five convolution blocks. Each block contains two convolution layers with the kernel size of $3 \times 3$ and the channel number of $128$. The stride of the first convolution layer is set to $2$ to achieve the downsampling. The decoder contains five bilinear upsampling layers, where each upsampling layer is followed by a convolution layer with the kernel size of $3 \times 3$ and the channel number of $128$. Each convolution layer is followed by a batch normalization layer and a leaky ReLU layer with a negative slope of $0.01$. The \emph{Decoder} is obtained by removing the encoder from the \emph{Encoder-Decoder}. Removing the upsampling layers from the \emph{Decoder} finally leads to the \emph{ConvNet}.}
    \label{exp_fig_architecture}
\end{figure*}
\begin{figure*}[t!]
\centering
\begin{subfigure}{1\textwidth}
\includegraphics[width=\textwidth]{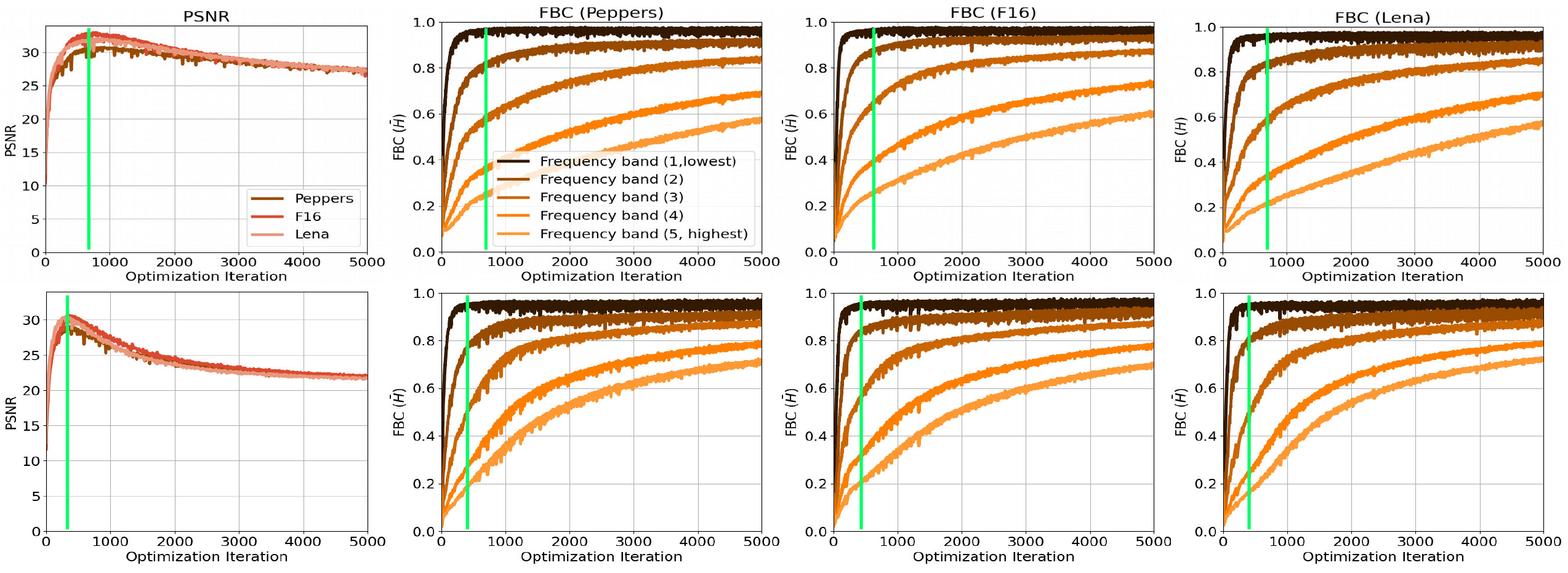}
\caption{\textbf{Image denoising (top: $\sigma{=}15$, bottom: $\sigma{=}25$)}}
\label{exp_fig_bias_task-a}
\end{subfigure}
\begin{subfigure}{1\textwidth}
\includegraphics[width=\textwidth]{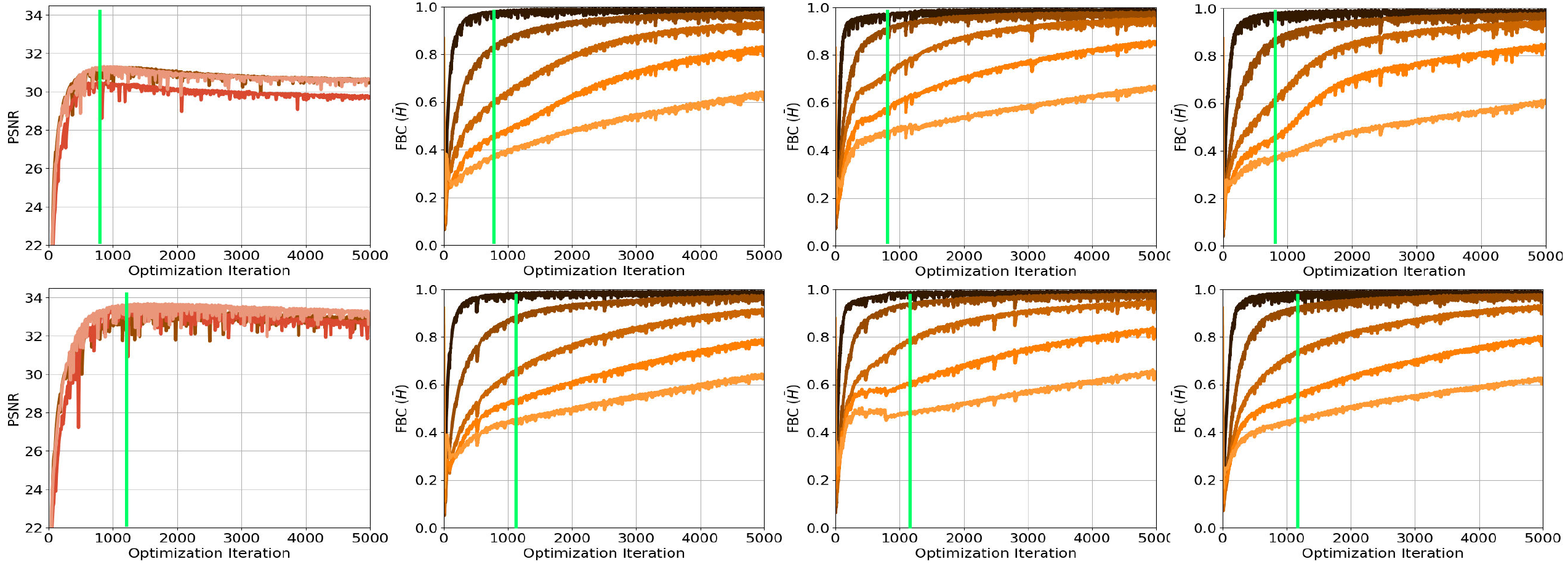}
\caption{\textbf{JPEG image deblocking (top: $quality{=}10$, bottom: $quality{=}20$)}}
\label{exp_fig_bias_task-b}
\end{subfigure}
\begin{subfigure}{1\textwidth}
\includegraphics[width=\textwidth]{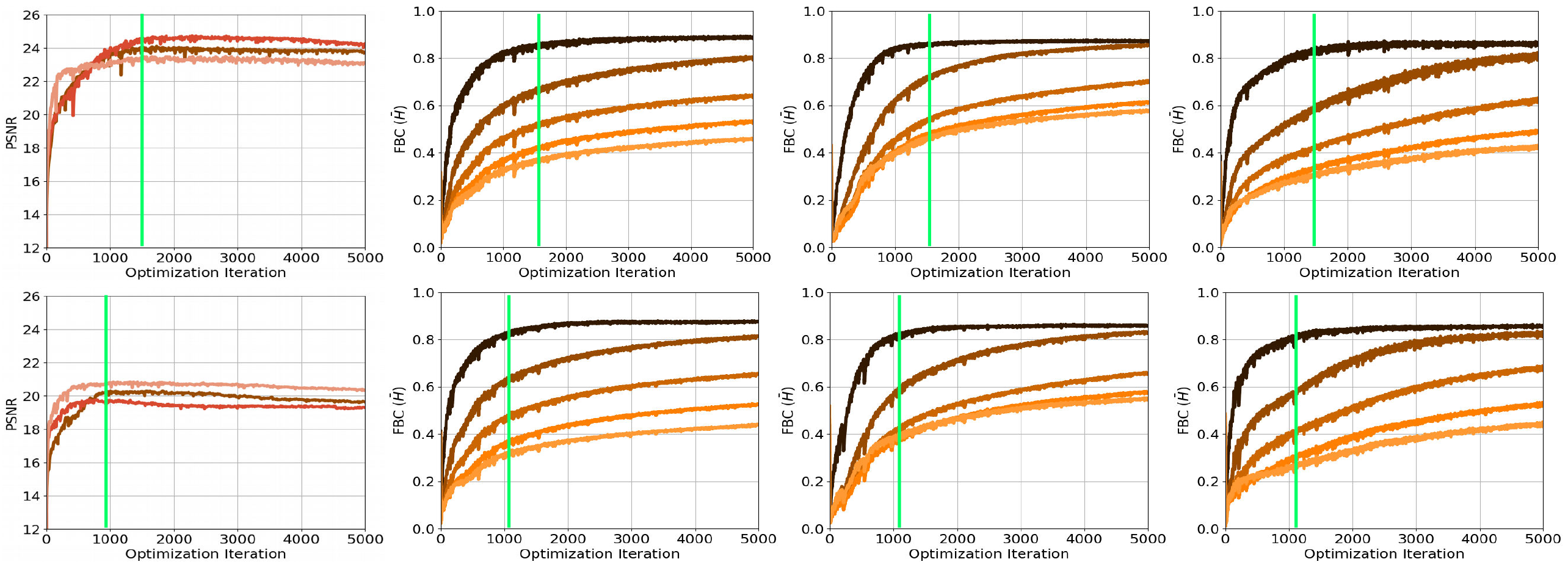}
\caption{\textbf{Image inpainting (top: $ratio{=}0.1$, bottom: $ratio{=}0.25$)}}
\label{exp_fig_bias_task-c}
\end{subfigure}

\caption{
\textbf{Spectral measurement of the deep image prior} on image denoising, JPEG image deblocking and image inpainting. The network of the deep image prior \citep{ulyanov2020deep} exhibits a spectral bias during optimization across inverse inaging problems, degradation levels and degraded images, where lower frequencies are learned faster and better than high-frequencies. The degraded images can be restored well when optimizations are stopped at the right time, as marked by the green vertical lines.}
\label{exp_fig_bias_task}    
\end{figure*}
\begin{figure*}[t!]
\centering
\begin{subfigure}{1\textwidth}
\includegraphics[width=\textwidth]{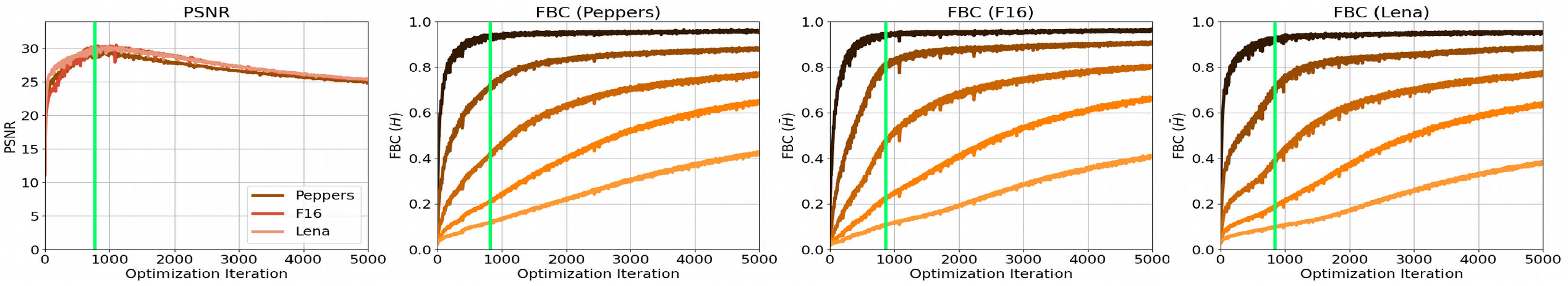}
\caption{\textbf{\emph{Decoder} architecture}}
\label{exp_fig_bias_net-a}
\end{subfigure}
\begin{subfigure}{1\textwidth}
\includegraphics[width=\textwidth]{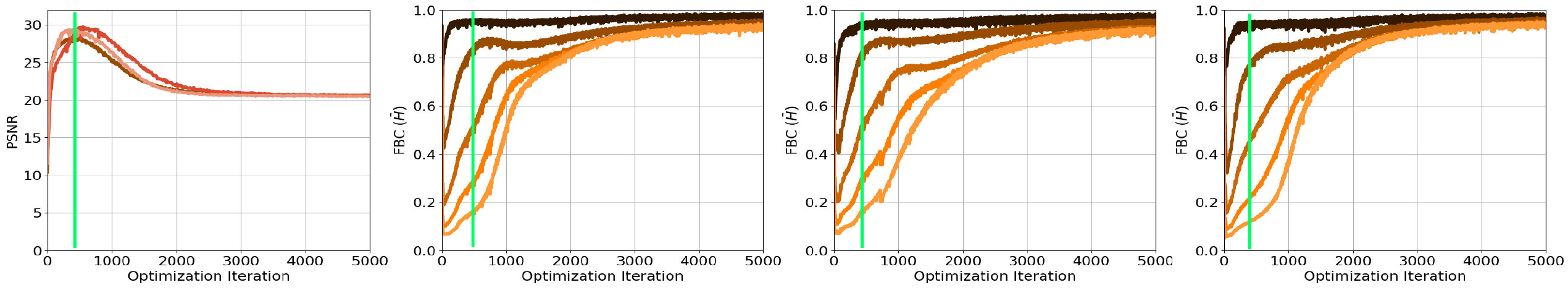}
\caption{\textbf{\textbf{\emph{ConvNet} architecture}}}
\label{exp_fig_bias_net-b}
\end{subfigure}
\caption{\textbf{Spectral measurement of the deep image prior} with different architectures on image denoising. The spectral bias is not specific to the \emph{Encoder-Decoder} architecture of \cite{ulyanov2020deep}. Alternative architectures, such as a \emph{Decoder} and a \emph{ConvNet}, also exhibit a bias towards specific image frequencies during optimization. Also, the \emph{ConvNet} learns higher frequencies faster than the \emph{Decoder} by removing the upsampling layers, but at the expense of reduced peak performance.}
\label{exp_fig_bias_net}    
\end{figure*}

\section{Measuring Spectral Bias} 
\label{sec_measure}
The literature attributes the ability of an untrained network to obtain restored results from degraded target images to a particular architecture, like an encoder-decoder, which imposes strong priors about natural images.
In this work, we show that the spectral bias leads the networks to capture deep image priors during optimization, beyond the choice of the architecture. We do so by introducing a metric, the Frequency Band Correspondence, which offers a spectral measurement of the deep image prior, revealing the low-frequency natural image signals are learned faster and better than high-frequency noise signals, and pinpoint why inverse images can be restored, when the network optimization is stopped at the right time.

\subsection{Frequency-Band Correspondence Metric}
\label{sec_fbc}
The proposed Frequency-Band Correspondence metric examines the input-output correspondence in the frequency domain across several frequency bands. For this metric, let $\{\theta^{(1)},\dots,\theta^{(T)}\}$ denote the trajectory of $T$ steps of gradient descent in the parameter space and let $\{f_{\theta^{(1)}},\dots,f_{\theta^{(T)}}\}$ denote the corresponding trajectory in the output space. We propose to analyze the Fourier spectrum of the output images $f_{\theta^{(t)},t{=}1,\dots,T}$ to show the convergence dynamics of different frequency components of the target image.
The Fourier spectrum of the output image $f_{\theta^{(t)}}$ is obtained by the Fourier transform $\mathcal{F}$, denoted as $\mathcal{F}\{f_{\theta^{(t)}}\}$ for step $t$. We similarly compute the Fourier transform for the target image $y_0$, denoted as $\mathcal{F}\{y_0\}$. We then compute an element-wise correspondence between both transforms as:
\begin{equation}
H_{\theta^{(t)}}=\frac{\mathcal{F}\{f_{\theta^{(t)}}\}}{\mathcal{F}\{y_0\}}.
\label{eq:fbc}
\end{equation}
Intuitively, $H_{\theta^{(t)}}$ denotes to what extent any deep image prior at step $t$ corresponds with image $y_0$ in the frequency domain; the closer the values are to 1, the higher the correspondence.
As we are interested in the spectral bias of the deep image prior, we divide the correspondence map into $N$ subgroups corresponding to $N$ non-overlapping frequency bands. Since the correspondence map is symmetrical around the center, we group it according to the distance between its elements and its center uniformly, as illustrated in Fig. \ref{fig:fbc}. To transform the 2D map to the 1D one, we compute the mean correspondence for each band, denoted as $\bar{H}_{\theta^{(t)}}^{(n)}$, with $n{=}1, \dots, N$. The value of $\bar{H}_{\theta^{(t)}}^{(n)}$ indicates the convergence dynamics of different frequency components of a target image. 

\subsection{Spectral Measurement of Deep Image Prior}
\label{sec_fbc_study}
We use this metric, denoted as FBC (Frequency-Band Correspondence), to measure how well the network output of the deep image prior corresponds to the target image as a function of $N$ frequency bands. Since the FBC metric is computed with the Fourier transform, our spectral measurement in this section denotes the frequency domain analysis. The Fourier transform $\mathcal{F}$ in Eq. (\ref{eq:fbc}) is implemented by means of the 2D Fast Fourier Transform, where only the magnitude is used to compute the Fourier spectrum of the images. We use $N{=}5$ where frequency bands are divided into the lowest frequency, low frequency, medium frequency, high frequency and the highest frequency. We perform empirical studies on three inverse imaging problems, including image denoising, JPEG image deblocking, and image inpainting with the `peppers', `F16' and `Lena', images from \cite{dabov2007image}. For image denoising, the image is degraded by adding Gaussian noise with two noise levels, including $\sigma{=}15$ and $\sigma{=}25$, following \cite{zhang2017beyond}. Following \cite{dong2015compression}, we evaluate JPEG image deblocking on the gray-scale images, which are compressed with the PIL encoder into two quality levels, including $quality{=}10$ and $quality{=}20$. For image inpainting, the image is degraded by using a central region mask, and we consider two hole-to-image area ratios, including $ratio{=}0.1$ and $ratio{=}0.25$, following \cite{pathak2016context}. Following \cite{ulyanov2018deep,ulyanov2020deep}, the network input is given as uniform noise between 0 and 0.1 with a depth of 32 by default. 

First, we investigate whether the network of the original deep image prior exhibits any form of spectral bias in its optimization. We take the \emph{Encoder-Decoder} architecture of \cite{ulyanov2018deep,ulyanov2020deep} and show its Frequency Band Correspondences for five frequency bands in Fig.~\ref{exp_fig_bias_task}. The plot highlights, across inverse imaging problems, degradation levels and degraded images, low frequencies are learned quickly and with high correspondence to the target image, while high frequencies are learned slower and with lower correspondence. We conclude that the network of the deep image prior during optimization has a spectral bias towards low frequencies, and this bias helps to obtain a meaningful performance. The peak PSNR (Peak Signal-to-Noise Ratio) performance of the deep image prior occurs when the lowest frequencies are matched nearly perfect, while the highest frequencies are less used, as marked by the green vertical lines. However, once the higher frequencies obtain a higher correspondence, the performance starts to drop.

Next, we show that such a spectral bias is not specific to the \emph{Encoder-Decoder} architecture. We take two other architectures as examples, as shown in Fig. \ref{exp_fig_architecture}. We remove the Encoder from the \emph{Encoder-Decoder} architecture of~\cite{ulyanov2018deep,ulyanov2020deep} to obtain the \emph{Decoder}. We additionally remove the upsampling layers from the \emph{Decoder} to get the \emph{ConvNet}.
Fig. \ref{exp_fig_bias_net-a} and \ref{exp_fig_bias_net-b} show that both \emph{Decoder} and \emph{ConvNet} learn low-frequency components of the target image faster than learning the high-frequency components, reaffirming the spectral bias. We also observe that \emph{ConvNet} learns high-frequency components faster than \emph{Decoder} by removing the upsampling layers, but at the expense of reduced peak performance.
Having established the architecture is not critical for the deep image prior, we use from now on the \emph{Decoder} as the default network architecture to benefit from a good trade-off between performance and run-time.

Our study provides a clear implication: untrained solutions for inverse imaging problems work by a latent ability to learn low frequencies faster than learning high frequencies. As natural images are well approximated by low-frequency components, degraded images can be restored well when optimizations are stopped at the right time. The network is optimized to fit the degraded image, in which higher frequencies consist of both structured high-frequency image details and random high-frequency noise. The structured high-frequency image details, that have self-similarity across the image, are fitted better and faster. However, once the random high-frequency noise is fitted over a certain level, which could affect the structured high-frequency image details, the output quality degrades. This behavior explains why the performance in the deep image prior degrades when training longer. Hence, a key enabler for improving the deep image prior is to control the spectral bias by restricting the fitting of random high-frequency noise in the output.
Our study also finds that the upsampling layer is beneficial for obtaining good peak performance, but may introduce too much spectral bias towards the low frequencies, slowing down the learning of desired high frequencies. Hence, it's a feasible way to balance peak performance and convergence by controlling the spectral bias in upsampling. 
\begin{figure*}[t]
\centering
\begin{subfigure}{1\textwidth}
\includegraphics[width=\textwidth]{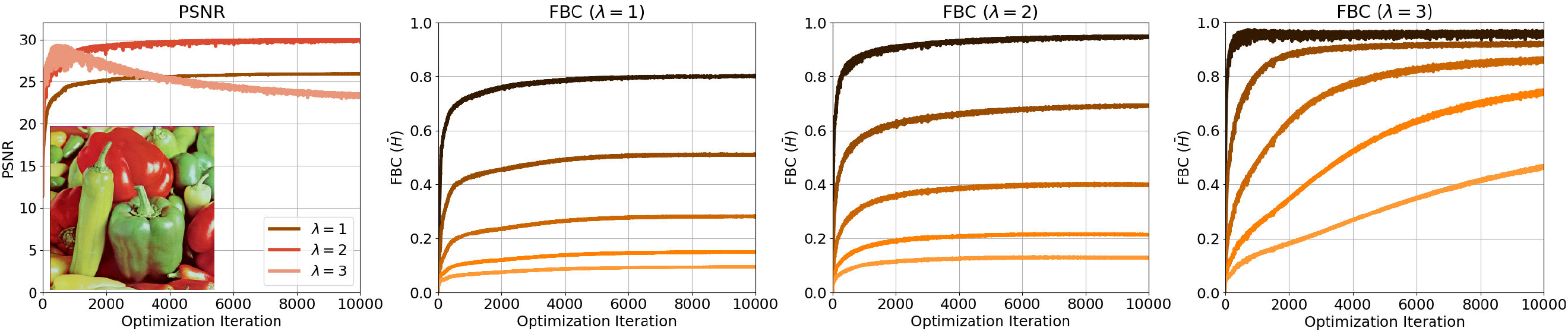}
\caption{\textbf{Peppers}}
\end{subfigure}
\begin{subfigure}{1\textwidth}
\includegraphics[width=\textwidth]{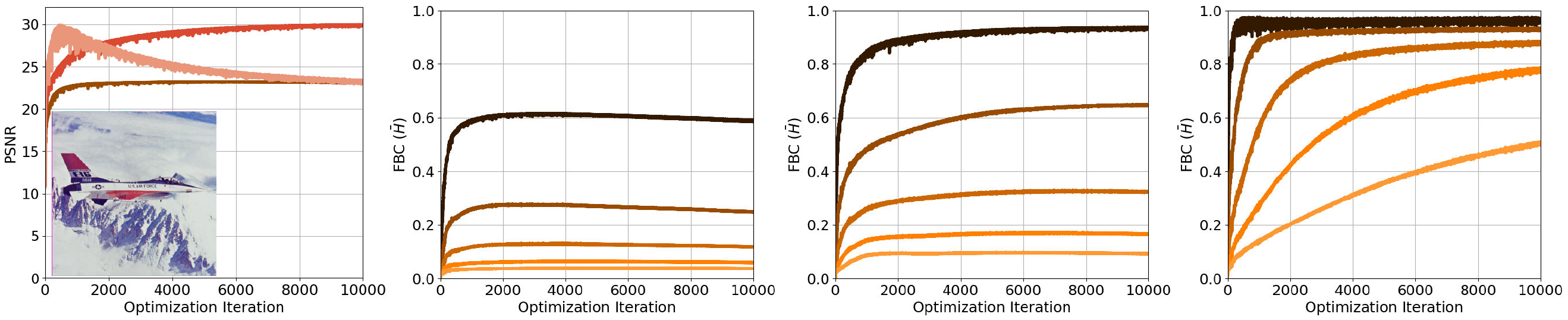}
\caption{\textbf{F16}}
\end{subfigure}
\begin{subfigure}{1\textwidth}
\includegraphics[width=\textwidth]{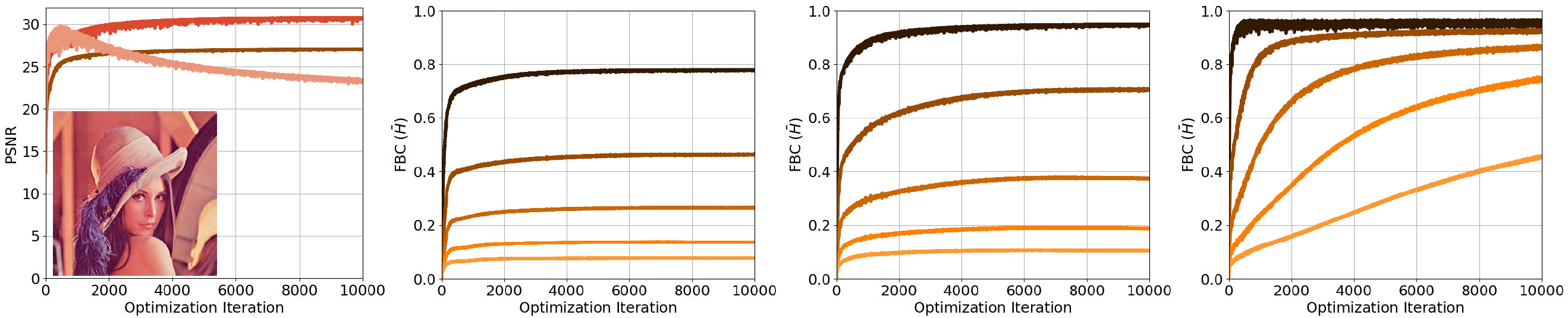}
\caption{\textbf{Lena}}
\end{subfigure}
\caption{
\textbf{Lipschitz-controlled spectral bias} for image denoising. Setting the right Lipschitz constant ($\lambda{=}2$) avoids performance decay while maintaining a high PSNR. Different constants result in different levels of spectral bias. A high constant ($\lambda{=}3$) still incorporates a lot of high-frequency noise signals, while a low constant ($\lambda{=}1$) fails to incorporate the important low frequency image signals. With the right balance ($\lambda{=}2$), we maintain the low frequencies while avoiding the high-frequency noise signals. 
}
\label{exp_fig_reg}    
\end{figure*}
\begin{figure*}[t]
\centering
\begin{subfigure}{1\textwidth}
\includegraphics[width=\textwidth]{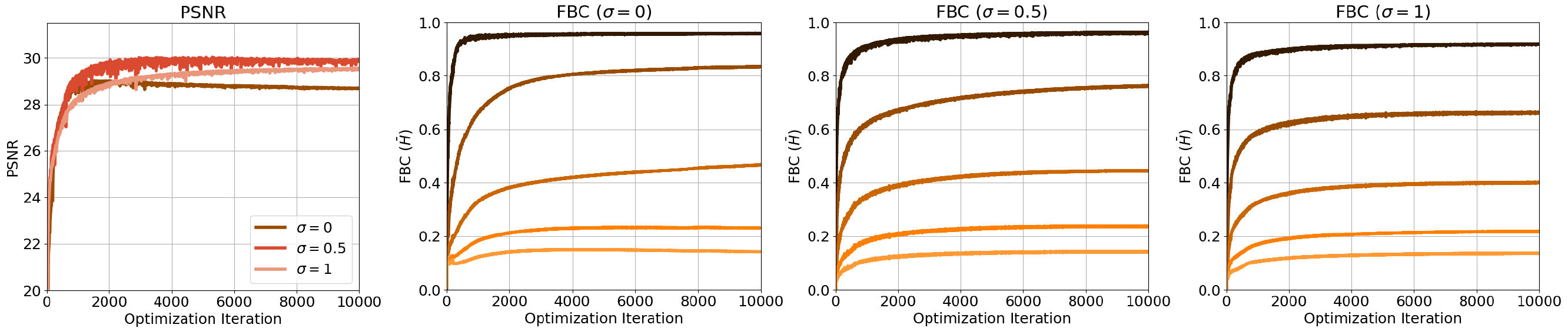}
\caption{\textbf{Peppers}}
\end{subfigure}
\begin{subfigure}{1\textwidth}
\includegraphics[width=\textwidth]{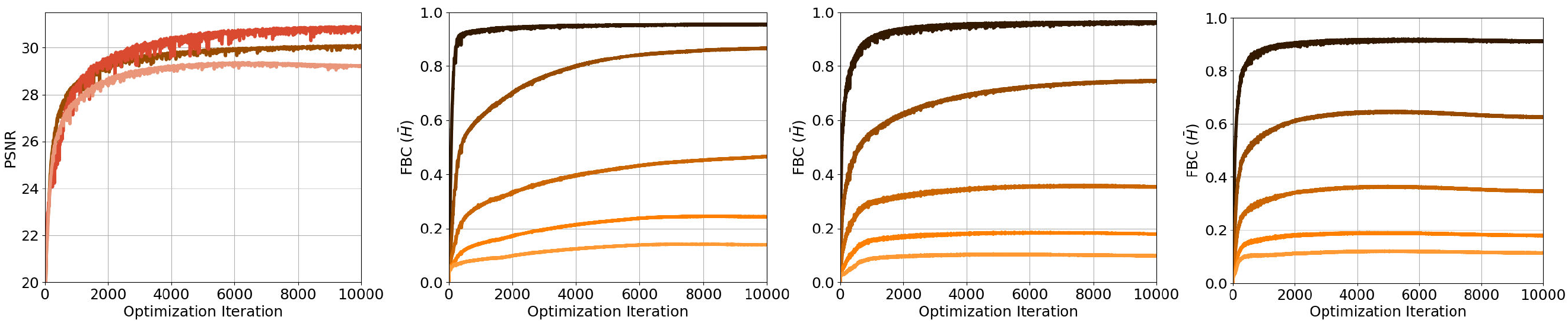}
\caption{\textbf{F16}}
\end{subfigure}
\begin{subfigure}{1\textwidth}
\includegraphics[width=\textwidth]{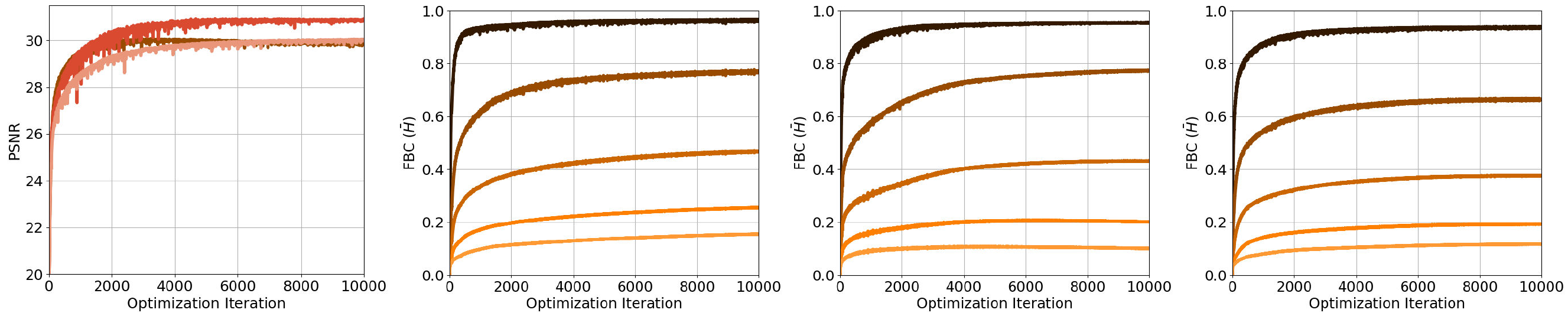}
\caption{\textbf{Lena}}
\end{subfigure}
\caption{
\textbf{Gaussian-controlled spectral bias} for image denoising. Varying the Gaussian kernel by $\sigma$ controls convergence and performance. Too small values ($\sigma{=}0$) results in worse performance, while too big values ($\sigma{=}1$) introduce too much smoothing, slowing down the convergence. With a suitable value 
($\sigma{=}0.5$), our upsampling introduces an appropriate spectral bias, leading to fast convergence and good denoising performance.}
\label{exp_fig_up}    
\end{figure*}
\begin{figure*}[!]
\centering
\begin{subfigure}{0.7\textwidth}
\includegraphics[width=\textwidth]{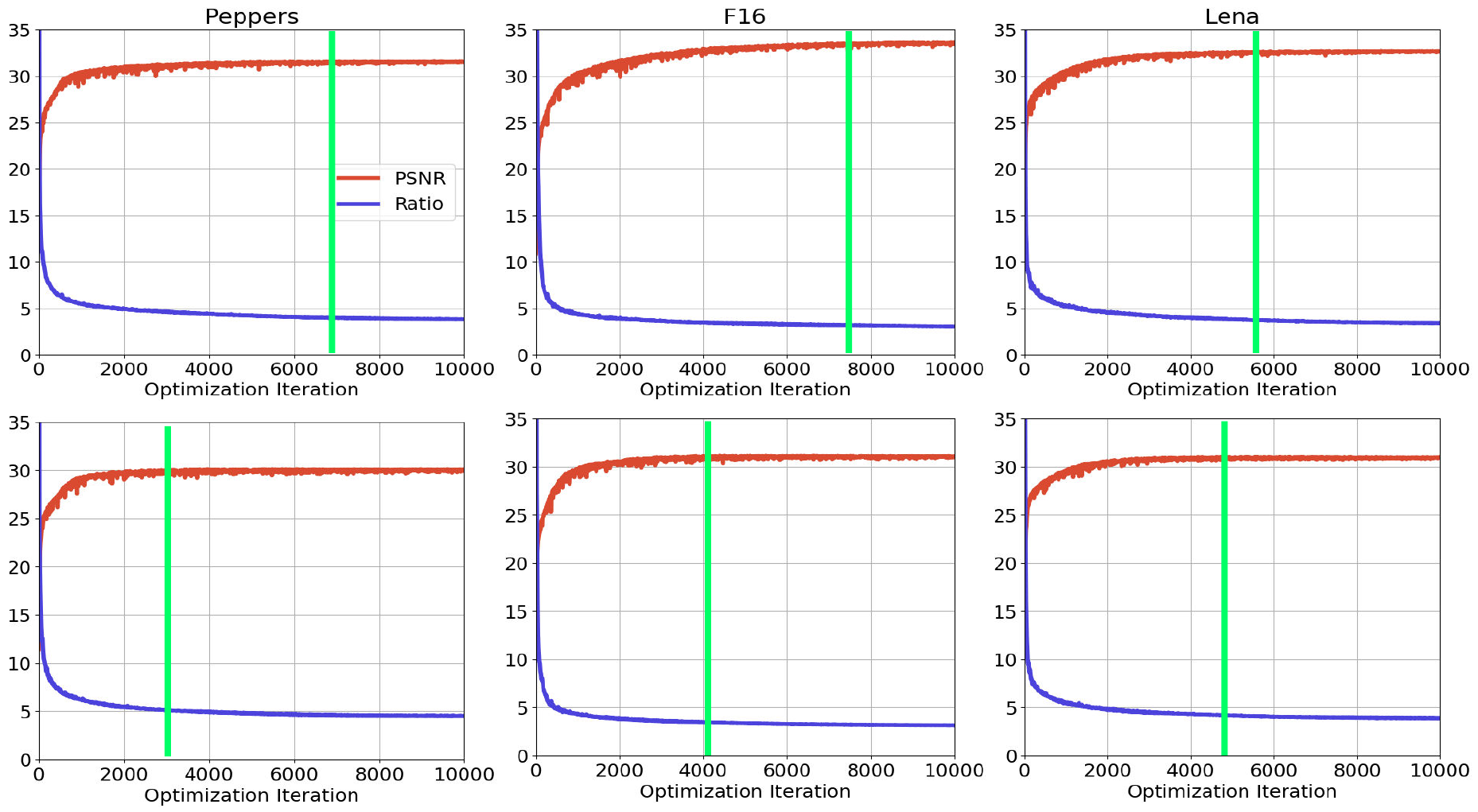}
\caption{\textbf{Image denoising (top: $\sigma{=}15$, bottom: $\sigma{=}25$)}}
\end{subfigure}
\begin{subfigure}{0.7\textwidth}
\includegraphics[width=\textwidth]{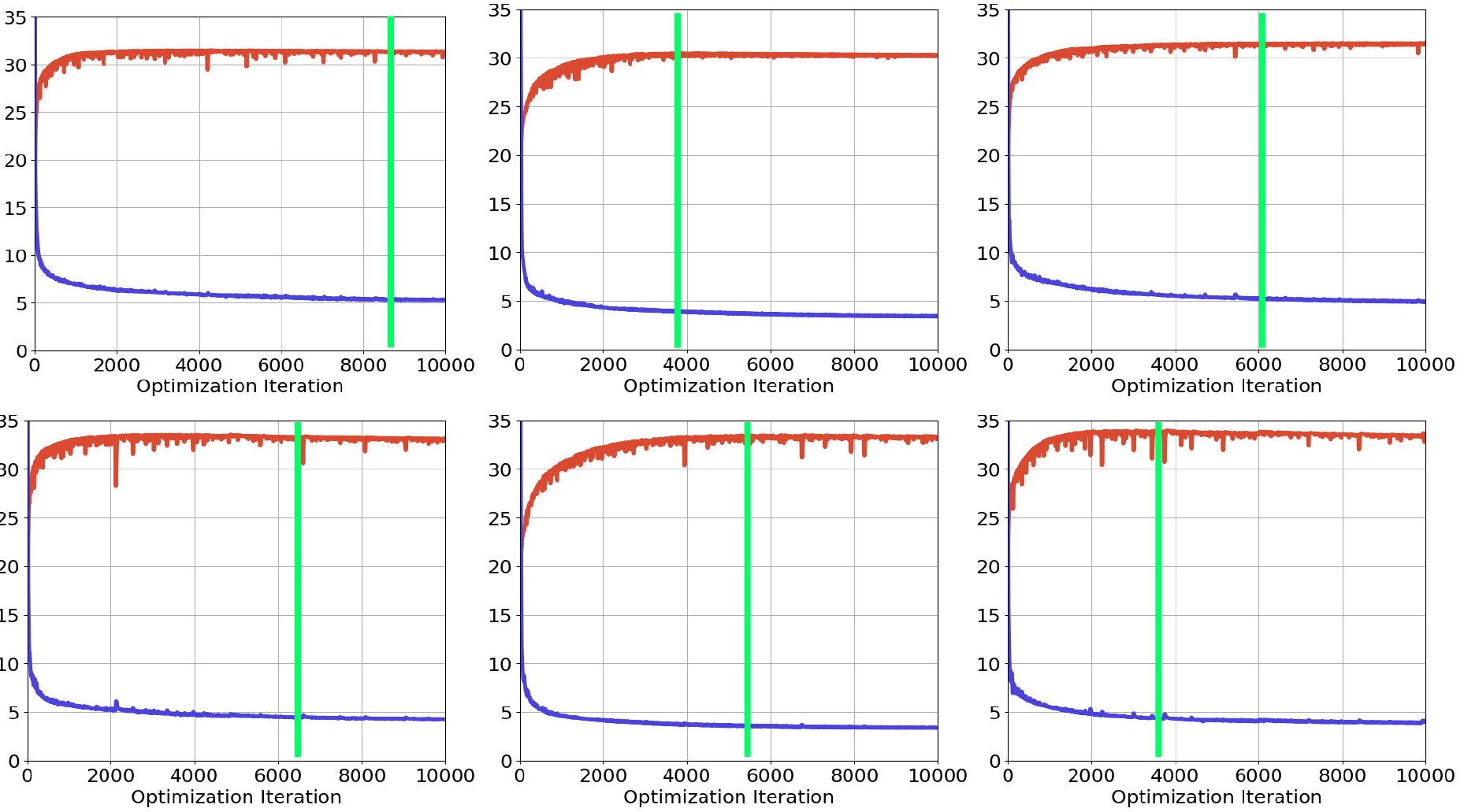}
\caption{\textbf{JPEG image deblocking (top: $quality{=}10$, bottom: $quality{=}20$)}}
\end{subfigure}
\begin{subfigure}{0.7\textwidth}
\includegraphics[width=\textwidth]{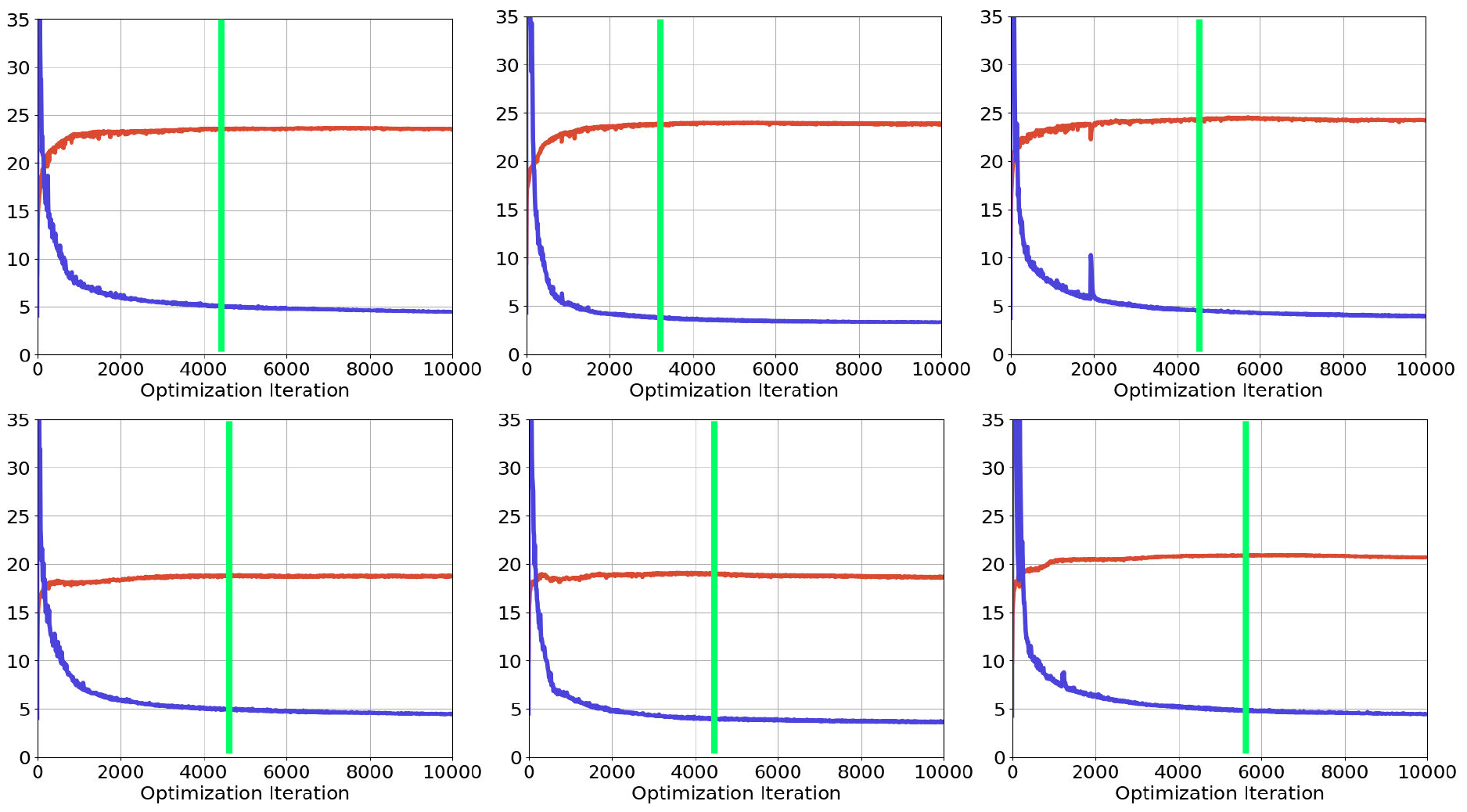}
\caption{\textbf{Image inpainting (top: $ratio{=}0.1$, bottom: $ratio{=}0.25$)}}
\end{subfigure}
\caption{\textbf{Automatic stopping criterion} evaluated on image denoising, JPEG image deblocking and image inpainting. The vertical green line shows the selected iteration by the proposed stopping criterion. Across inverse imaging problems, degradation levels and degraded images, we observe the optimization can be stopped earlier, with a minimal performance loss compared to a fixed stop at 10,000 iterations.}
\label{exp_fig_stop}    
\end{figure*}
%
\section{Controlling Spectral Bias}
\label{sec_control}
We exploit the measured spectral bias to avoid the degradation of performance over iterations and to balance peak performance and convergence.
We do so by controlling spectral biases in the two core layer types of inverse imaging networks: the convolution layer and the upsampling layer. We present a Lipschitz-controlled approach for the convolution and a Gaussian-controlled approach for the upsampling layer. The approaches are general in their setup, making them applicable to any network form and scale. Besides these two methods for controlling spectral bias, we further introduce a simple stopping criterion to avoid superfluous computation.

\subsection{Lipschitz-Controlled Spectral Bias}
From the point of view of the frequency domain, the Fourier spectrum of the network indicates its ability to learn higher frequencies. 
Lower frequencies are learned first, while higher frequencies are learned later in the optimization process. This implies that the ability of the network to learn higher frequencies is gradually enhanced by optimizing the learnable layers. Improving the Fourier spectrum of the network is only achievable through adjusting the spectrum of the learnable layers. 
Based on this observation, we aim to upper bound the Fourier coefficients of the convolutional layers, for the sake of constraining the Fourier spectrum of the network.
We are able to impose an upper bound on the Fourier coefficients of a convolution layer by enforcing Lipschitz continuity, according to \cite{katznelson2004introduction}. Specifically, if a convolution layer $f$ is Lipschitz continuous, there exists a constant $L$ for any inputs $x,y$ satisfying $\lVert f(x)-f(y)\rVert \le L \lVert x-y \rVert$. The minimum over all such values satisfying this condition is called the Lipschitz constant of $f$, denoted by $C$. Then the Fourier coefficients of $f$, \ie, $|\hat{f}(\boldsymbol{k})|$, is bounded by,
\begin{equation}
    |\hat{f}(\boldsymbol{k})| \le \frac{C}{|\boldsymbol{k}|^2}.
\end{equation}
Further, the Lipschitz constant of a convolution layer is bounded by the spectral norm of its parameters. Then we obtain,
\begin{equation}
    |\hat{f}(\boldsymbol{k})| \le \frac{C}{|\boldsymbol{k}|^2} \le \frac{\lVert \boldsymbol{w} \rVert_{sn}}{|\boldsymbol{k}|^2},
\end{equation}
where $\boldsymbol{w}$ is the weight
of a convolution layer $f$, and $\lVert \cdot \rVert_{sn}$ denotes the spectral norm, which can be approximated relatively quickly using a few iterations of the power method \citep{miyato2018spectral}. The power law $|\boldsymbol{k}|^{-2}$ indicates that the spectral decay is stronger towards higher frequencies, which means that learning higher frequencies requires a higher spectral norm. Thus, we are able to manipulate the ability of a convolution layer in learning higher frequencies by upper bounding its spectral norm to a specific value $\lambda$ with $\frac{\boldsymbol{w}} {\max(1, \lVert \boldsymbol{w} \rVert_{sn}/\lambda)}$. Where we leave the weight matrix $\boldsymbol{w}$ untouched if its spectral norm is lower than $\lambda$. Otherwise, we normalize $\boldsymbol{w}$ by $\lVert \boldsymbol{w} \rVert_{sn}/\lambda$.

To accelerate and stabilize the optimization, batch normalization \citep{ioffe2015batch} is often used after convolution layers. However, we find it is not compatible with our Lipschitz constraining as its output is invariant to the channel weight vector norm $\lVert \boldsymbol{w} \rVert_{p}$, \ie, 
\begin{equation}
  BN(\boldsymbol{w}\boldsymbol{x}/\lVert \boldsymbol{w} \rVert_{p}) = BN(\boldsymbol{w}\boldsymbol{x}),
\end{equation}
where $\boldsymbol{x}$ denotes the channel input.
We therefore propose a Lipschitz normalization by exploring the idea of combining Lipschitz constraining with a special version of batch normalization: mean-only batch normalization. We only subtract out the minibatch means, without dividing by the minibatch standard deviations. The Lipschitz normalization is defined as:
\begin{equation}
\label{eq_lipnorm}
\text{LN}(\boldsymbol{w},\boldsymbol{x}) = \ \frac{\boldsymbol{w}\boldsymbol{x}}{\max(1, \lVert \boldsymbol{w} \rVert_{sn}/\lambda)}-\mu + b,  
\end{equation}
where $\mu$ denotes the channel mean of the pre-activation $\boldsymbol{w}\boldsymbol{x}$ and $b$ is a scalar bias term. The Lipschitz normalization layer is inserted between a convolutional layer and a ReLU activation. With this normalization, the Lipschitz constant of a convolution layer is bounded by the hyperparameter $\lambda$. As a result, we can manipulate the ability of the network in learning high frequencies by tuning $\lambda$, leading to a controlled spectral bias of the deep image prior.

\subsection{Gaussian-Controlled Spectral Bias}
\label{sec_upsampling}
Upsampling is an important operation in network architectures for inverse imaging problems, as it produces high-resolution outputs from low-resolution inputs. Well-known approaches such as the bilinear and nearest neighbor upsampling have a constant smoothing effect \citep{chakrabarty2019spectral,heckel2019denoising}. Different tasks, however, might operate best under different levels of smoothing. Too strong a smoothing introduces too much spectral bias towards lower frequencies. This slows down the learning of the desired higher frequencies, delaying convergence of optimization (as shown in Fig. \ref{exp_fig_bias_net}). Therefore, we propose an upsampling method which allows controlling the amount of smoothing and is capable of balancing performance and convergence.

We first decompose the upsampler into an expansion and a filtering step. Let $x_i$ be the i-th channel of input $x$. For expansion, $x_i$ is padded with a ``bed of nails” scheme, \ie, inserting $s-1$ zeros between the pixels of $x_i$ along its rows and columns. Such a ``bed of nails” expansion creates a high-frequency replica of the original signal. To smooth out the noisy high-frequencies, we perform filtering by convolving the upsampled signal with an interpolating filter. We use a Gaussian filter sampled by $\mathcal{N}(0, \sigma^2)$.
Hence, we define our Gaussian upsampling by:
\begin{equation}
\label{eq_upGaussian}
    \text{Up}(x_i) = \uparrow_s(x_i)*G_\sigma,
\end{equation}
where $\uparrow_s(x_i)$ denotes expanding $x_i$ with factor $s$, $*$ is the convolution operation, $G_\sigma$ denotes the Gaussian filter. In the frequency domain, we obtain the Fourier spectrum of our upsampling by,
\begin{equation}
    \mathcal{F}(\text{Up}(x_i)) = \mathcal{F}(\uparrow_s(x_i)) \odot \mathcal{F}(G_\sigma),
\end{equation}
where $\mathcal{F}$ denotes the Fourier transform, $\odot$ is the Hadamard product and $\mathcal{F}(G_\sigma)[k]{=}1/e^{2\pi ^2 \sigma^2 k^2}$. We manipulate the Fourier spectrum of our upsampling by choosing different $\sigma$, allowing us to control the spectral bias in the upsampling. 
\subsection{Automatic Stopping Criterion}
With the ability to control the spectral bias, we can fix the number of iterations for network optimization without fear of performance degradation. As different tasks have different levels of convergence, however, using a fixed number of iterations still leads to redundant optimization. To improve efficiency, we introduce a simple criterion to automatically perform early stopping.

It is well known that an image looks blurry when there is a high amount of low frequencies in its Fourier spectrum. We exploit this property by computing the blurriness and sharpness for an output image and use their ratio as the metric to stop the optimization. In case of a spectral bias, low frequencies will be learned first, while high-frequencies will be learned later. Our Lipschitz normalization limits the ability of the network in learning high frequencies to an upper bound. Hence, when this upper bound is reached, the ratio of blurriness to sharpness of the output image will converge as well. To that end, we design the following measure:
\begin{equation}
\begin{split}
     r({f_\theta}) = & \mathcal{B}({f_\theta}) / \mathcal{S}({f_\theta}),\\
     \Delta r({f_\theta}^t) = & \bigg|\frac{1}{n}\sum_{i=1}^n r\left({f_\theta}^{(t-i)}\right) - \frac{1}{n}\sum_{i=1}^n r\left({f_\theta}^{(t-n-i)}\right) \bigg|,
\end{split}
\end{equation}
where ${f_\theta}$ denotes the output image and ${f_\theta}^{(t)}$ denotes an instance in iteration $t$. $\mathcal{B}({f_\theta})$ denotes the blurriness of the output image $y$ computed using \cite{crete2007blur}. $\mathcal{S}({f_\theta})$ denotes the sharpness of the output image $y$ computed using \cite{bahrami2014fast}. $r({f_\theta})$ denotes the ratio of blurriness to sharpness of the output image ${f_\theta}$. Then, $\frac{1}{n}\sum_{i=1}^n r\left({f_\theta}^{(t-i)}\right)$ computes the mean ratio of output images from iteration $(t-1)$ to $(t-n)$, and $\frac{1}{n}\sum_{i=1}^n r\left({f_\theta}^{(t-n-i)}\right)$ computes the mean ratio of output images from iteration $(t-n-1)$ to $(t-2n)$. If their absolute difference is smaller than a constant value $\epsilon$, the optimization is stopped.

Compared to the ratio $r$ itself, the ratio difference $\Delta r$ between optimization iterations is independent of the images. Since the deep image prior no longer suffers from performance degradation with the controlled spectral bias, the ratio $r$ barely changes when the performance is stable. Thus, we can set the ratio difference threshold $\epsilon$ to a small value, like 0.01. As the main benefit of the auto-stopping is to avoid redundant computation, it does not directly affect the inverse imaging performance. Note that the stopping criterion fails for the original deep image prior \citep{ulyanov2018deep,ulyanov2020deep} because the high-frequency components of its output image keeps increasing until the degraded target image is fully fitted.
\subsection{Performance Analysis}
\label{sec_control_study}
We empirically analyze the deep image prior with the Lipschitz-controlled spectral bias, the Gaussian-controlled spectral bias and the automatic stopping criterion.

\textbf{Lipschitz-controlled spectral bias.}
Following the work of \cite{ulyanov2018deep,ulyanov2020deep}, we use bilinear upsampling in this experiment. 
In Eq. (\ref{eq_lipnorm}), $\lambda$ is the only parameter which controls the ability of the network in learning high frequencies. Finding the best $\lambda$ for each image is still an open question. Here we just empirically study three settings, \ie,~$\lambda{=}1$,$\lambda{=}2$, and $\lambda{=}3$.
The spectral norm $\lVert \boldsymbol{w} \rVert_{sn}$ is estimated with the power iteration method \citep{miyato2018spectral}. The results are shown in Fig.~\ref{exp_fig_reg}. Setting a suitable constraint (\eg, $\lambda{=}2$) results in a PSNR curve without performance decay. The FBC graphs show this is because setting a low Lipschitz constant amplifies the spectral bias. High frequencies are hardly incorporated at all, while low frequencies still obtain a high correspondence to the target image.
Using a too high constraint (\eg, $\lambda{=}3$) results in a similar performance peak and decay as the original deep image prior.
When using a too low constraint (\eg, $\lambda{=}1$), we not only suppress high frequencies, but also the low frequencies, which generally corresponds to the structure of the image, hampering the performance.
We conclude, utilizing Lipschitz normalization with a suitable value of $\lambda$ addresses the problem of performance degradation.

\textbf{Gaussian-controlled spectral bias.}
Next, we study the effect of the Gaussian-controlled spectral bias to balance performance and convergence. We replace the bilinear upsampling with our Gaussian upsampling and use $\lambda{=}2$ to maintain the effect of the Lipschitz-controlled spectral bias on avoiding performance degradation. 
We consider Gaussian upsampling with three settings in Eq. (\ref{eq_upGaussian}), $\sigma{=}0$, $\sigma{=}0.5$, $\sigma{=}1$ where the kernel size is fixed to $5 \times 5$. We show the effect of different settings on the denoising performance and amount of spectral bias in Fig.~\ref{exp_fig_up}.
The smaller the value for $\sigma$, the faster the convergence is reached. However, a too small value \eg, $\sigma{=}0$ results in worse performance, because the upsampling reduces to the ``bed of nails'' expansion.
A value of $\sigma{=}1$ introduces too much smoothing, slowing down the convergence. With a suitable value, \eg, $\sigma{=}0.5$, our upsampling introduces an appropriate spectral bias, leading to fast convergence and good denoising performance. 
Furthermore, compared to the widely used upsampling, like bilinear upsampling (refer to its performance in Fig. \ref{exp_fig_reg}), our upsampling achieves a better trade-off between performance and convergence. 
We conclude our upsampling allows to control the spectral bias, enabling us to improve the performance of deep image prior for inverse imaging problems like image denoising.

\textbf{Stopping criterion.}
Finally, we analyze the effect of the proposed stopping criterion on image denoising, JPEG image deblocking and image inpainting. For each problem, we evaluate on different degradation levels, as specified before in Section \ref{sec_fbc_study}. We use $n{=}100$ and $\epsilon{=}0.01$ throughout the experiment. We set the fixed stopping iteration to 10,000. We show the dynamics of the Peak Signal-to-Noise score and ratio values in Fig. \ref{exp_fig_stop}. We observe the stopping criterion is effective, it reduces the number of required iterations considerably with only a minimal loss in performance, across inverse imaging problems, degradation levels, and degraded images. For the worst performing ``F16'' image for denoising with $\sigma{=}25$, the PSNR drops from 31.04 to 30.98 when reducing the iterations from 10,000 to 3,896. We also found that the performance in terms of PSNR changes less than $0.1$ when the ratio difference threshold $\epsilon$ ranges from $0.001$ to $0.1$. A bigger threshold means the optimization stopped earlier.

%% file: 5-experiment.tex
\begin{figure*}[t!]
\centering
\includegraphics[width=0.99\linewidth]{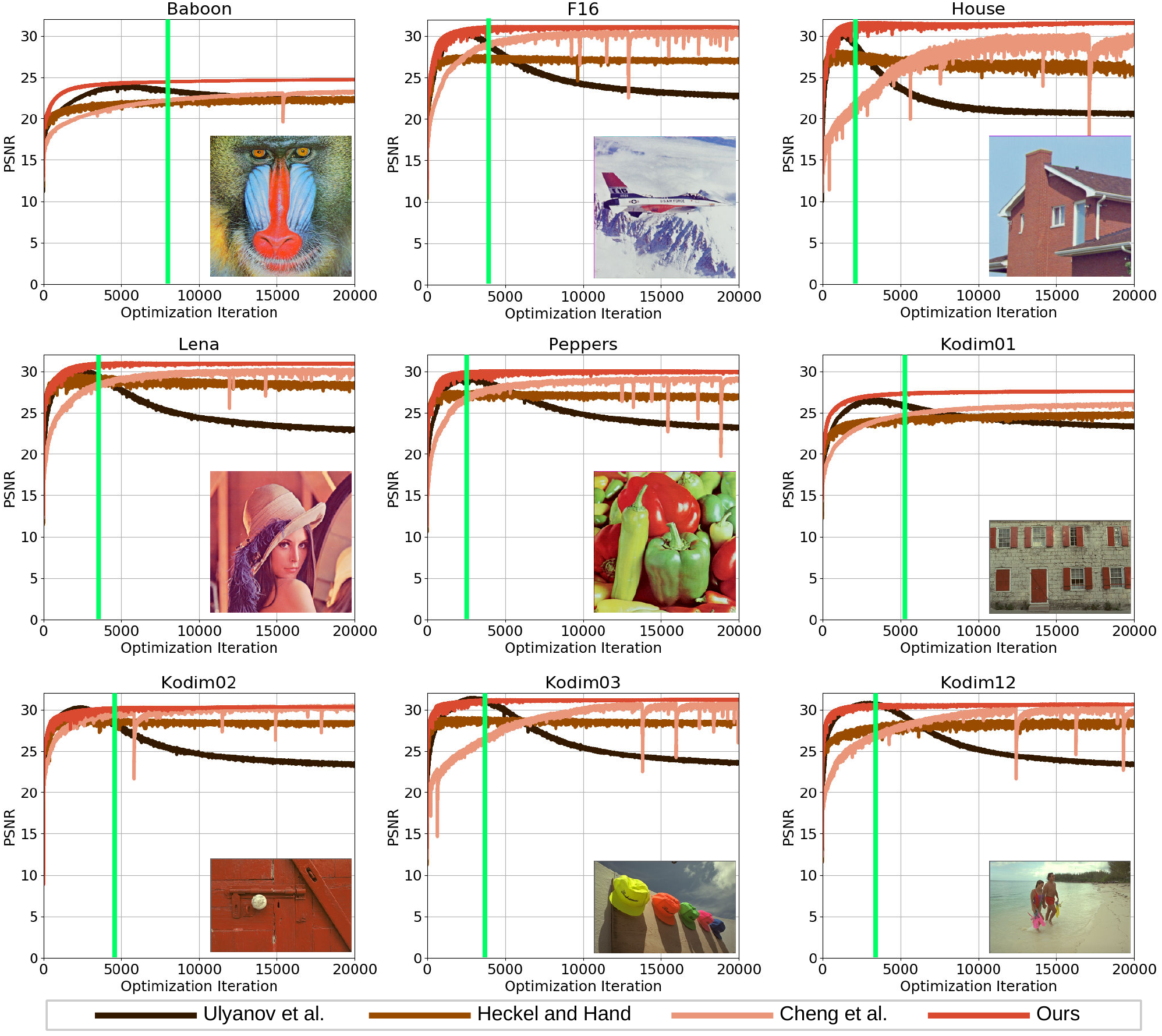}
\caption{\textbf{Image denoising.} PSNR scores of various methods over multiple iterations for removing additive Gaussian white noise with $\sigma {=} 25$. Compared to \citep{ulyanov2020deep}, our method doesn't suffer from performance degradation, and we can stop the optimization automatically at an appropriate moment for each image (marked by the green vertical lines), leading to good PSNR scores. Compared to \cite{heckel2018deep} and \cite{cheng2019bayesian}, we either achieve a faster convergence or obtain a higher PSNR score.}
\label{fig:exp-2-1}
\end{figure*}
\begin{figure*}[t!]
\centering
\includegraphics[width=0.99\linewidth]{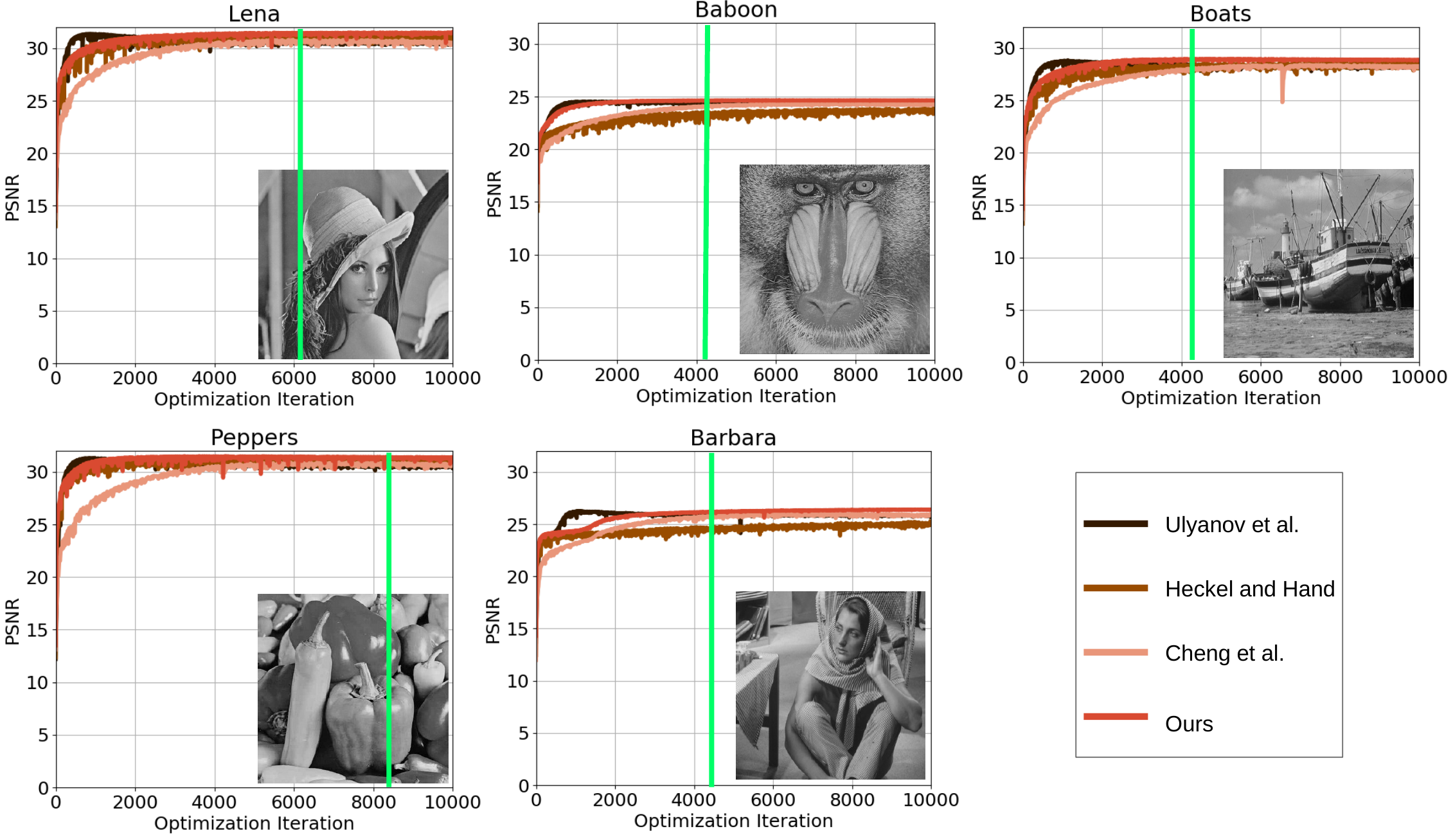}
\caption{\textbf{JPEG image deblocking.} PSNR scores of various methods when reducing artifacts of a compressed JPEG image with $quality {=} 10$. We again observe that the performance of the deep image prior \citep{ulyanov2020deep} degrades. \cite{cheng2019bayesian} and \cite{heckel2018deep} do not suffer from degradation, at the expense of either reduced performance or slow convergence. Our method achieves a good trade-off between PSNR score and convergence (marked by the green vertical lines).}
\label{fig:exp-2-2}
\end{figure*}
\begin{figure*}[t!]
\centering
\includegraphics[width=0.99\linewidth]{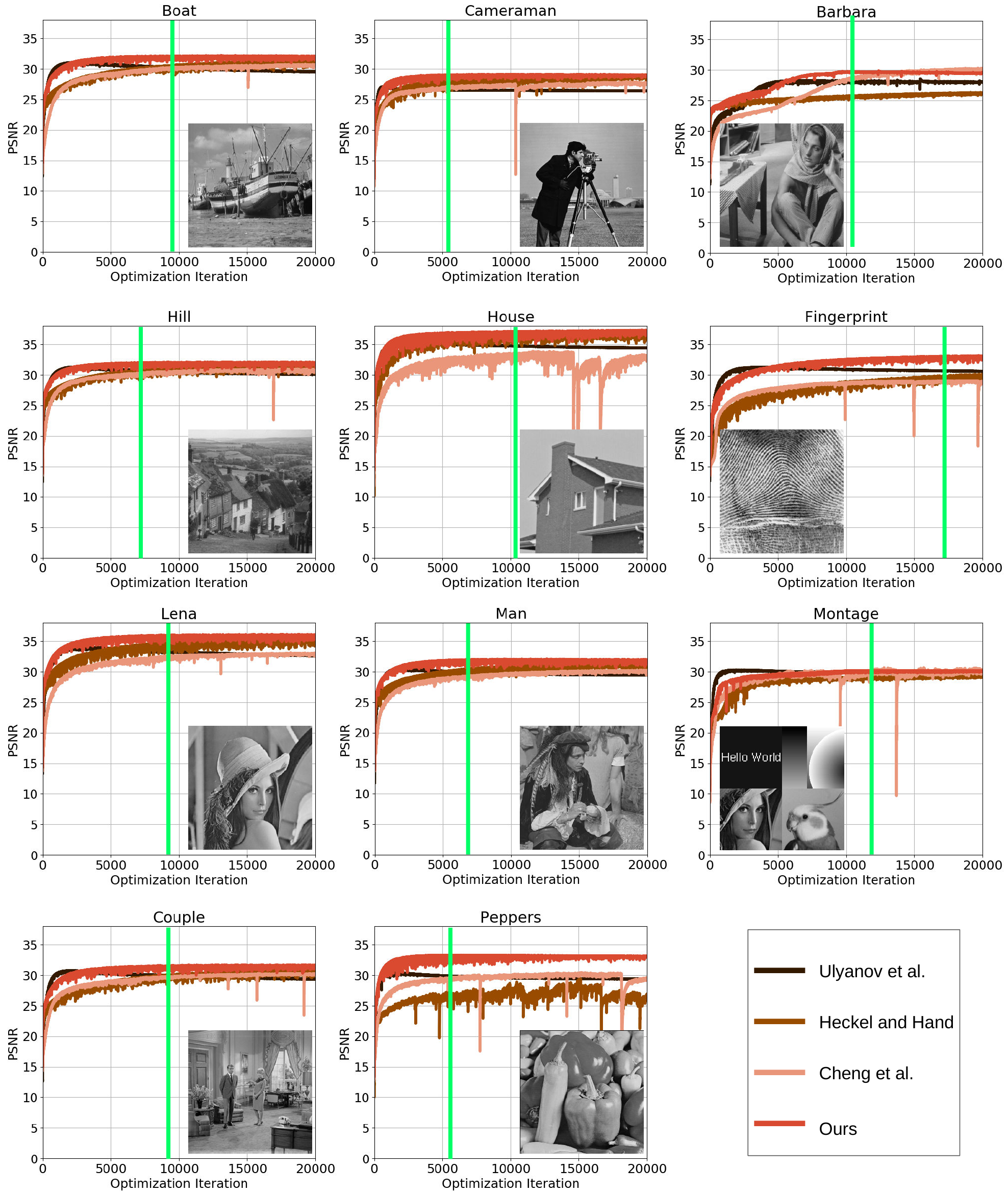}
\caption{\textbf{Image inpainting.} PSNR scores of various methods for pixel inpainting. We again observe the degradation of performance over iterations for the deep image prior \citep{ulyanov2020deep}. \cite{cheng2019bayesian} and \cite{heckel2018deep} do not suffer from the degradation problem, at the expense of either reduced performance or slow convergence. Our method achieves a good trade-off between PSNR score and convergence (marked by the green vertical lines). }
\label{fig:exp-2-3}
\end{figure*}
%
\begin{figure*}[t!]
\centering
\begin{subfigure}{0.204\textwidth}
\includegraphics[width=\textwidth]{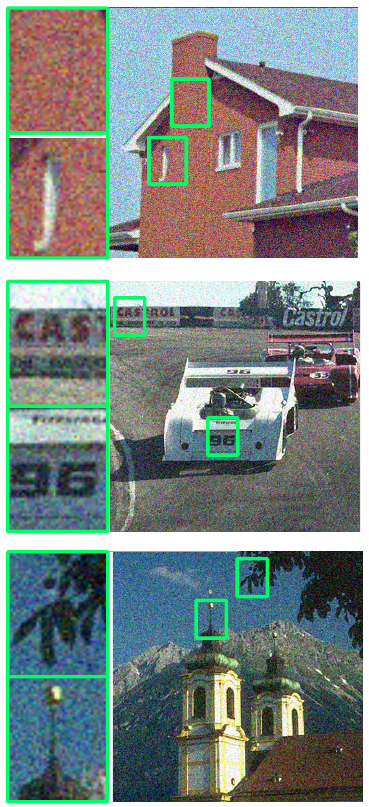}
\caption{\textbf{Noisy image}}
\end{subfigure}
\begin{subfigure}{0.204\textwidth}
\includegraphics[width=\textwidth]{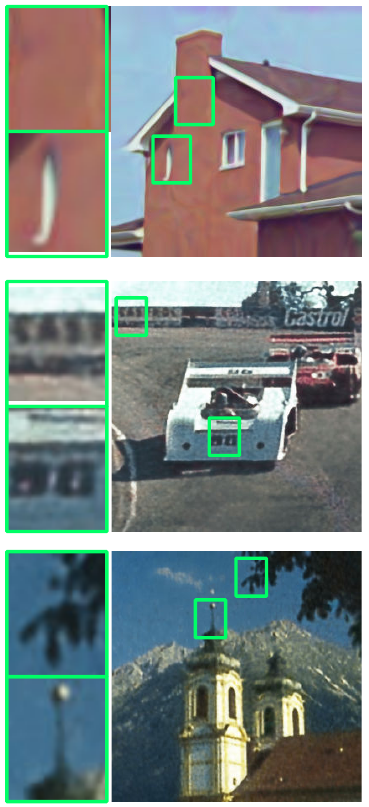}
\caption{\textbf{Heckel and Hand}}
\end{subfigure}
\begin{subfigure}{0.204\textwidth}
\includegraphics[width=\textwidth]{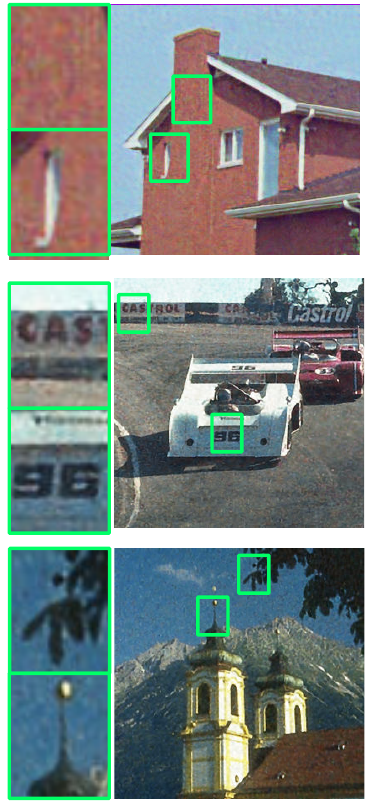}
\caption{\textbf{Cheng \etal}}
\end{subfigure}
\begin{subfigure}{0.204\textwidth}
\includegraphics[width=\textwidth]{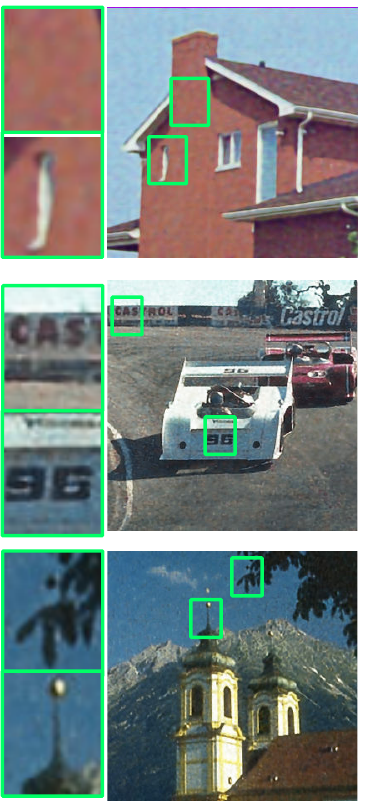}
\caption{\textbf{Ours}}
\end{subfigure}
\caption{\textbf{Image denoising}. The goal is to remove the additive Gaussian white noise with $\sigma {=} 25$. From the top regions masked by the green rectangles, we observe the method of \cite{cheng2019bayesian} still overfits some high-frequency noise, while our method does not. From the bottom regions masked by the green rectangles, we observe the method of \cite{heckel2018deep} has difficulty preserving high-frequency edges, while our method performs better.}
\label{exp_fig_denoise}    
\end{figure*}
%
\begin{figure*}[t!]
\centering
\begin{subfigure}{0.204\textwidth}
\includegraphics[width=\textwidth]{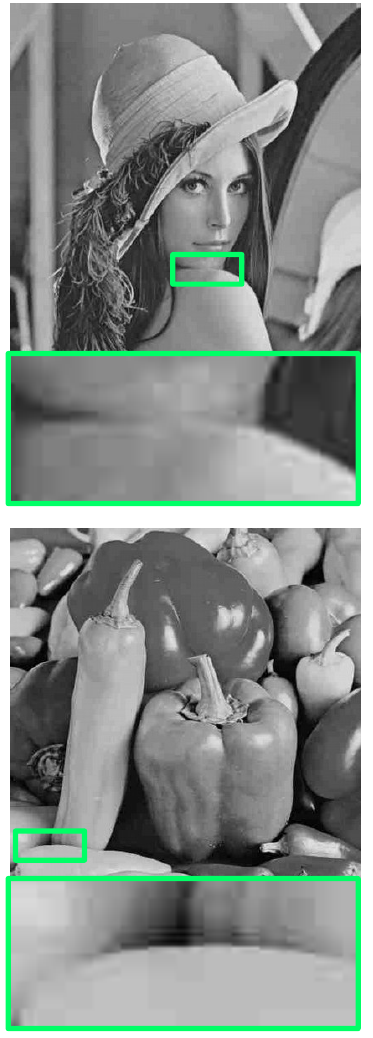}
\caption{\textbf{Corrupted image}}
\end{subfigure}
\begin{subfigure}{0.198\textwidth}
\includegraphics[width=\textwidth]{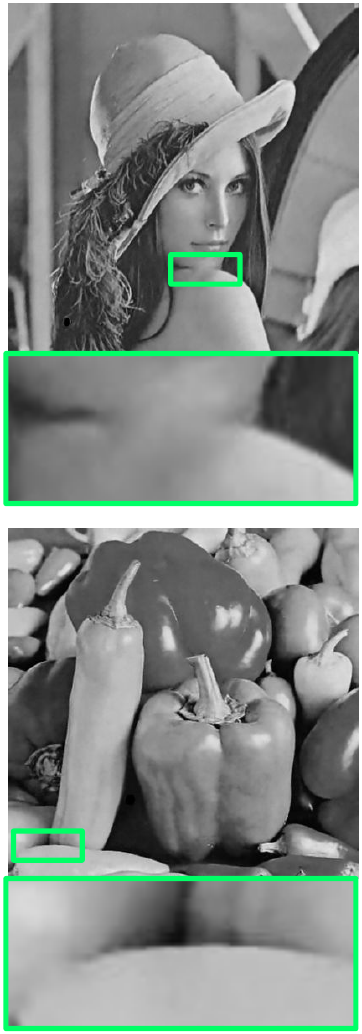}
\caption{\textbf{Cheng et al.}}
\end{subfigure}
\begin{subfigure}{0.20\textwidth}
\includegraphics[width=\textwidth]{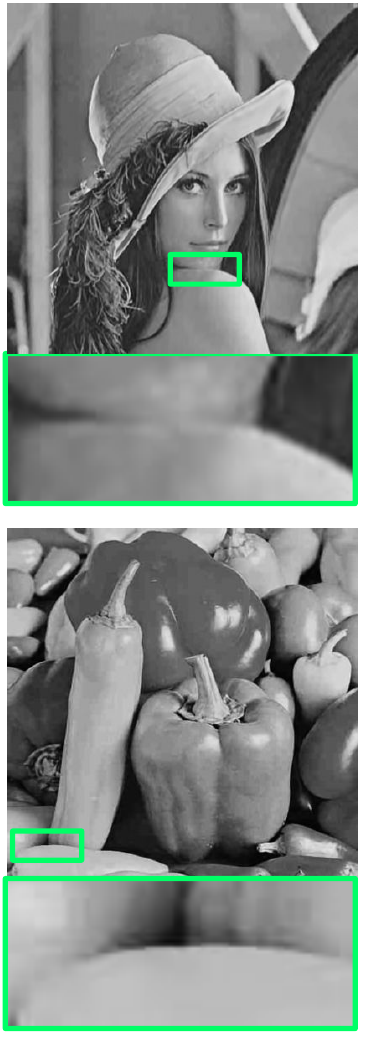}
\caption{\textbf{\textbf{Ours}}}
\end{subfigure}
\begin{subfigure}{0.206\textwidth}
\includegraphics[width=\textwidth]{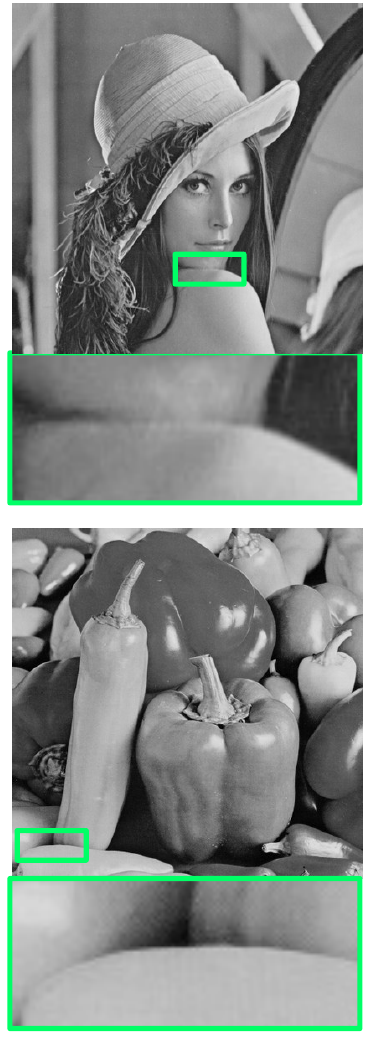}
\caption{\textbf{Ground truth}}
\end{subfigure}
\caption{\textbf{JPEG image deblocking}. The goal is to reduce the artifacts of the compressed JPEG image with $quality {=} 20$. From the regions masked by the green rectangles, we observe our method performs well, especially when reducing the artifacts and recovering high-frequency image details.}
\label{exp_fig_deblocking}    
\end{figure*}
%
%
\begin{figure*}[t!]
\centering
\begin{subfigure}{0.205\textwidth}
\includegraphics[width=\textwidth]{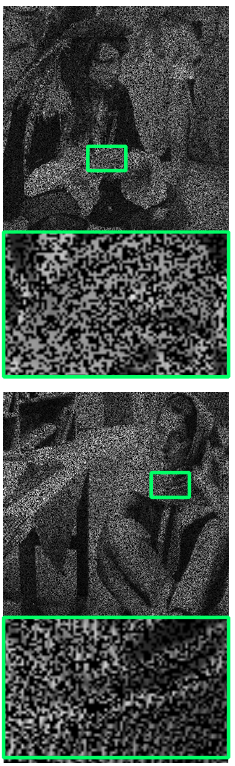}
\caption{\textbf{Corrupted image}}
\end{subfigure}
\begin{subfigure}{0.207\textwidth}
\includegraphics[width=\textwidth]{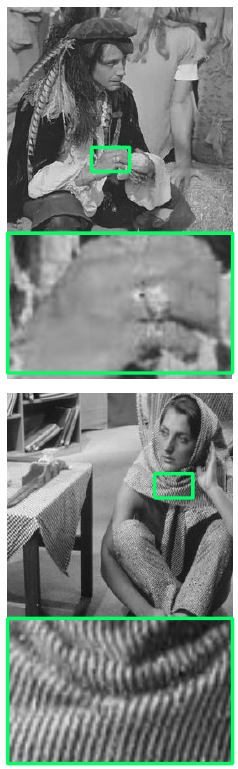}
\caption{\textbf{Cheng et al.}}
\end{subfigure}
\begin{subfigure}{0.205\textwidth}
\includegraphics[width=\textwidth]{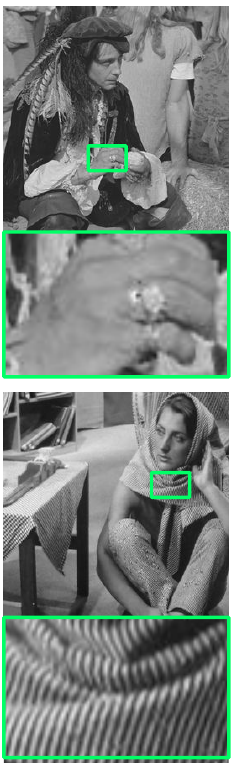}
\caption{\textbf{\textbf{Ours}}}
\end{subfigure}
\begin{subfigure}{0.208\textwidth}
\includegraphics[width=\textwidth]{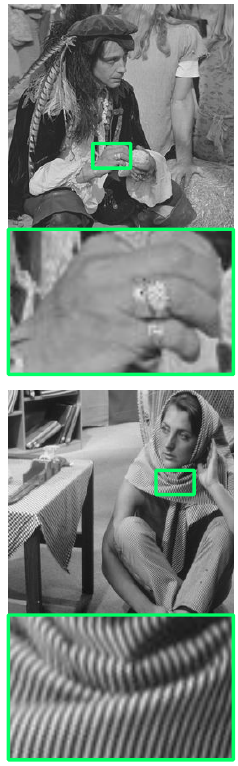}
\caption{\textbf{Ground truth}}
\end{subfigure}
\caption{\textbf{Image inpainting}. The goal is to reconstruct the $50\%$ missing pixels resulting from a binary Bernoulli mask. From the regions masked by the green rectangles, we observe our method performs well, especially when recovering high-frequency details.}
\label{exp_fig_inpainting}    
\end{figure*}
%
%
\begin{figure*}[t!]
\centering
\begin{subfigure}{0.208\textwidth}
\includegraphics[width=\textwidth]{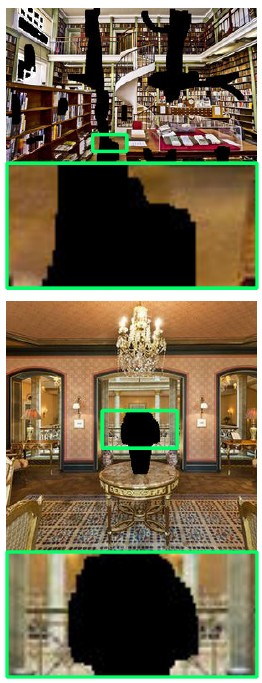}
\caption{\textbf{Corrupted image}}
\end{subfigure}
\begin{subfigure}{0.214\textwidth}
\includegraphics[width=\textwidth]{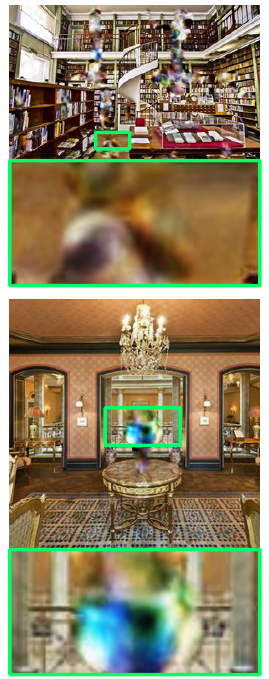}
\caption{\textbf{Heckel and Hand}}
\end{subfigure}
\begin{subfigure}{0.206\textwidth}
\includegraphics[width=\textwidth]{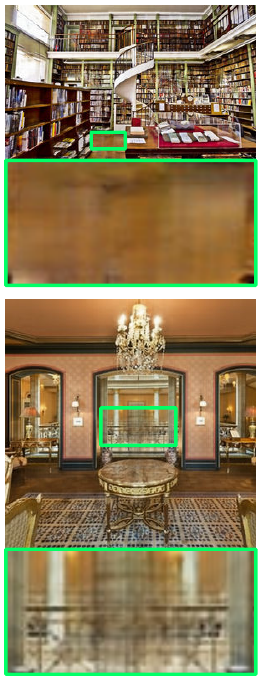}
\caption{\textbf{Cheng et al.}}
\end{subfigure}
\begin{subfigure}{0.209\textwidth}
\includegraphics[width=\textwidth]{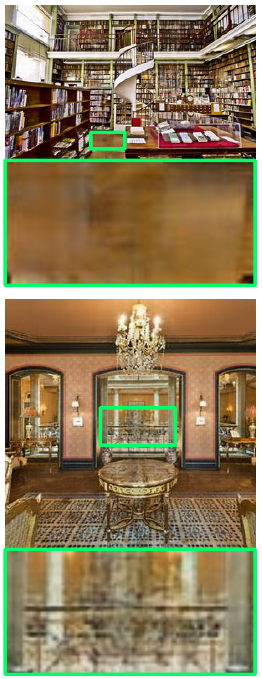}
\caption{\textbf{\textbf{Ours}}}
\end{subfigure}
\caption{\textbf{Image inpainting}. The goal is to reconstruct the missing pixels resulting from a binary region mask. From the regions masked by the green rectangles, we observe our method performs better than \cite{heckel2018deep} and as good as \cite{cheng2019bayesian}.}
\label{exp_fig_hole_inpainting}    
\end{figure*}
\begin{figure*}[t!]
\centering
\begin{subfigure}{0.206\textwidth}
\includegraphics[width=\textwidth]{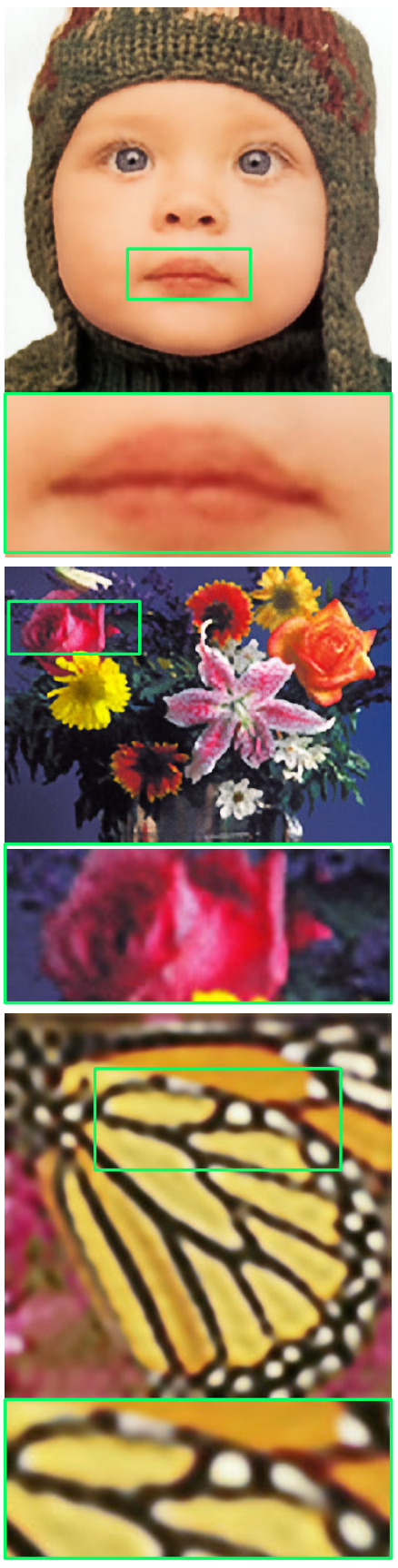}
\caption{\textbf{Ulyanov et al.}}
\end{subfigure}
\begin{subfigure}{0.209\textwidth}
\includegraphics[width=\textwidth]{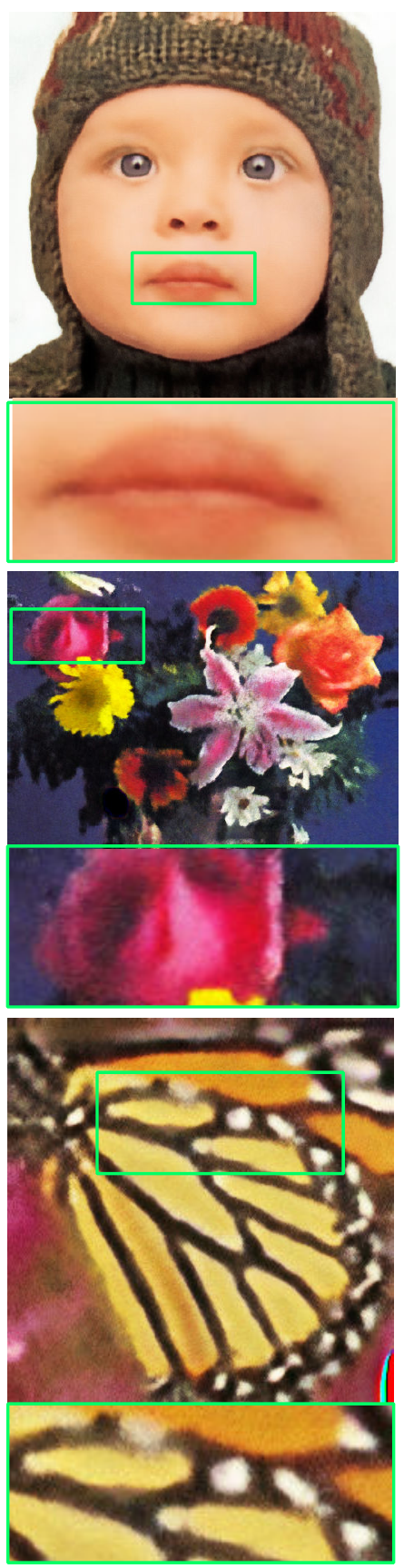}
\caption{\textbf{Cheng et al.}}
\end{subfigure}
\begin{subfigure}{0.206\textwidth}
\includegraphics[width=\textwidth]{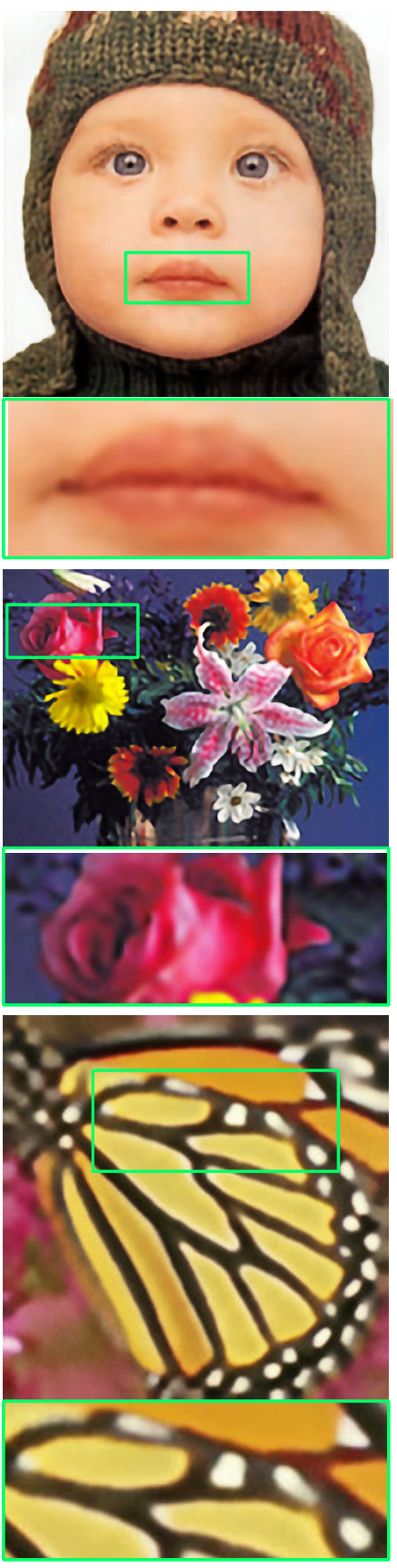}
\caption{\textbf{Ours}}
\end{subfigure}
\begin{subfigure}{0.202\textwidth}
\includegraphics[width=\textwidth]{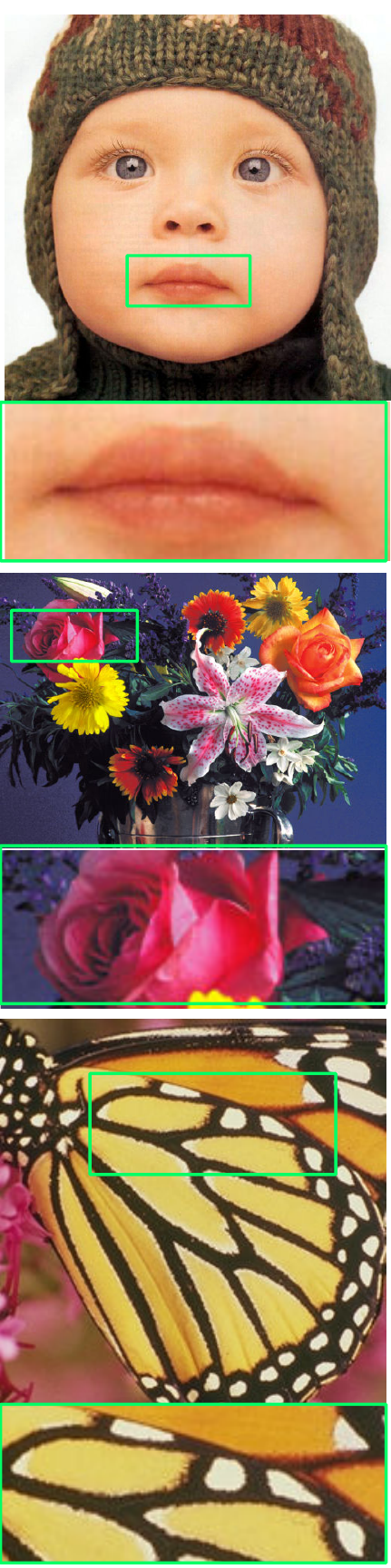}
\caption{\textbf{Ground truth}}
\end{subfigure}
\caption{\textbf{Super-resolution}. Results on the `baby' image and the `flowers' image for $4 \times$ super-resolution, and on the `butterfly' image for $8 \times$ super-resolution. From the regions masked by the green rectangles, we observe our method is able to better recover details with fewer artifacts (best viewed digitally).}
\label{exp_fig_sr}    
\end{figure*}
%
\begin{figure*}[t!]
\centering
\begin{subfigure}{0.206\textwidth}
\includegraphics[width=\textwidth]{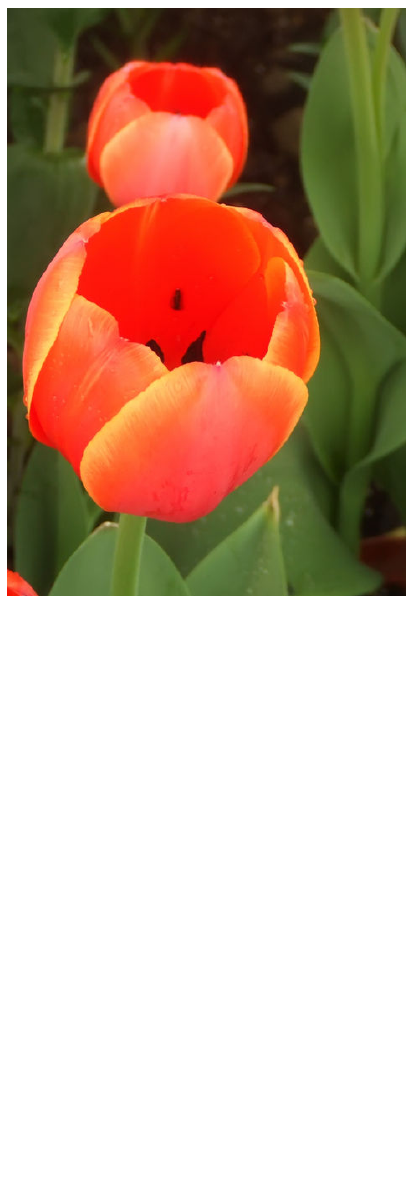}
\caption{\textbf{Original image}}
\end{subfigure}
\begin{subfigure}{0.203\textwidth}
\includegraphics[width=\textwidth]{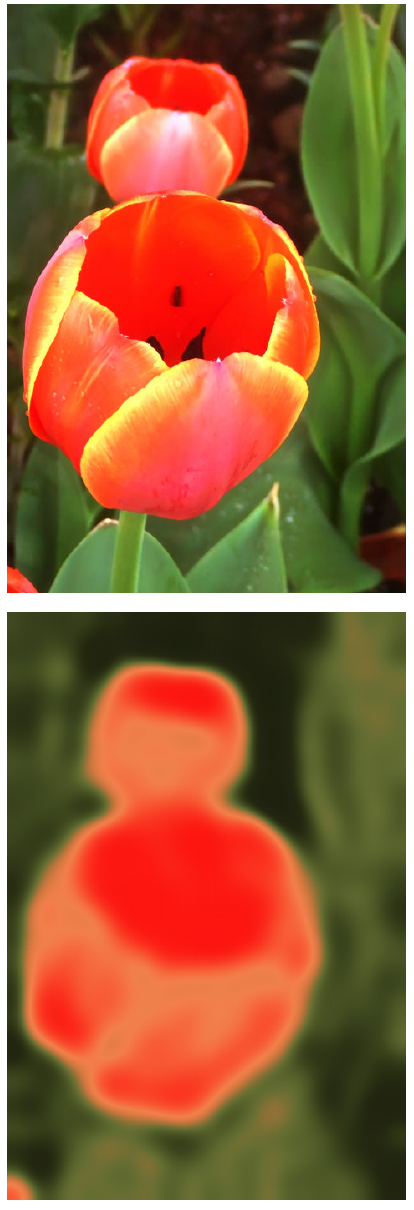}
\caption{\textbf{$\lambda {=} 1$}}
\end{subfigure}
\begin{subfigure}{0.205\textwidth}
\includegraphics[width=\textwidth]{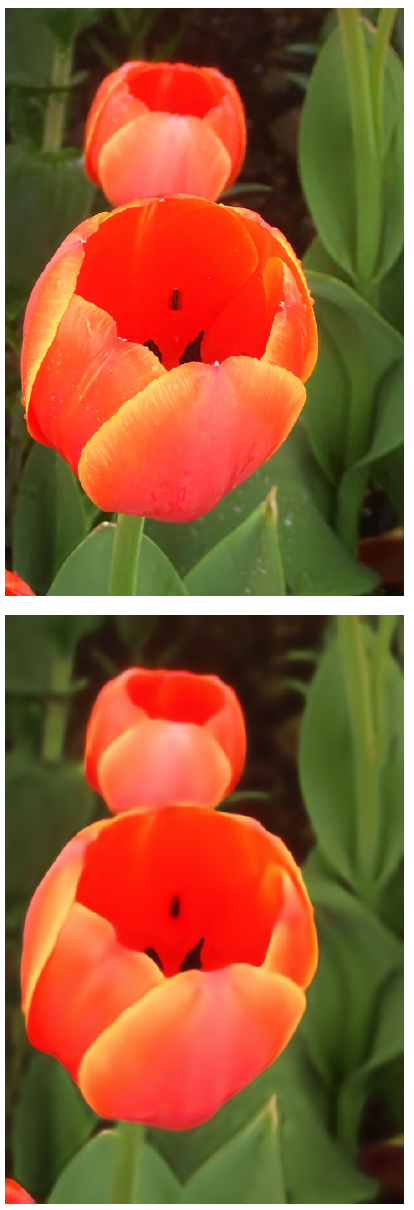}
\caption{\textbf{$\lambda {=} 2$}}
\end{subfigure}
\begin{subfigure}{0.200\textwidth}
\includegraphics[width=\textwidth]{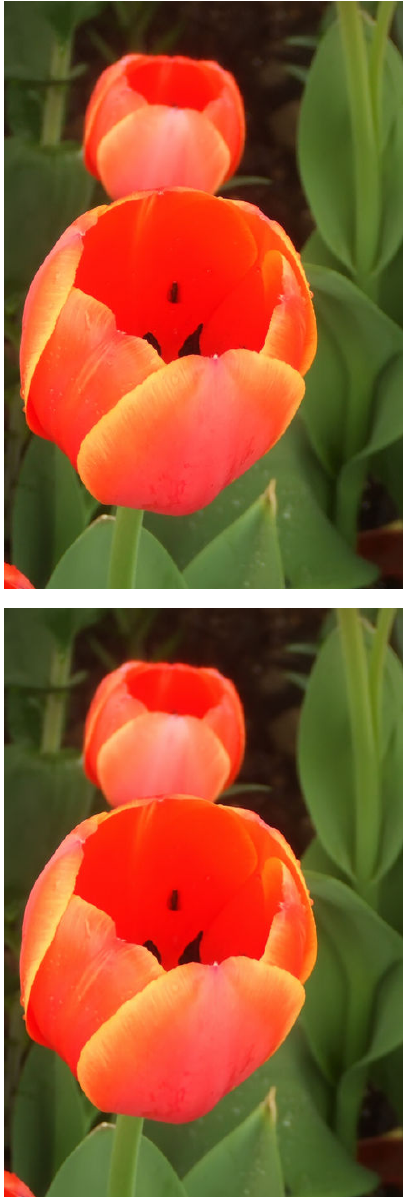}
\caption{\textbf{$\lambda {=} 3$}}
\end{subfigure}
\caption{\textbf{Image enhancement}. The goal is to enhance the image details. We obtain the smoothed images (second row) using the controlled deep image priors with different $\lambda$, as defined in Eq. (\ref{eq_lipnorm}). We then subtract the smoothed version from the original image to get fine details and enhance them (first row). The smaller the $\lambda$, the higher smoothness of the output images and the more enhancement to the image details.}
\label{exp_fig_enhancement}    
\end{figure*}
%

\section{Applications}
\label{exp_sota}
With the gained ability to control the spectral bias in the deep image prior, we consider four inverse imaging applications and one image enhancement application for comparative evaluation: image denoising, JPEG image deblocking, image inpainting, image super-resolution and image detail enhancement. On all tasks, we compare to the deep image priors of \cite{ulyanov2018deep,ulyanov2020deep}, \cite{heckel2018deep} and \cite{cheng2019bayesian}. For reference, we also report the results obtained by classical methods like \citep{dabov2007image}, and supervised-learning based methods like \citep{zhang2017beyond}.

We report our results with the \emph{Decoder}, introduced in Section \ref{sec_fbc_study}, as our network architecture. Lipschitz normalization with $\lambda{=}2$ and Gaussian upsampling with $\sigma{=}0.5$ are combined into the \emph{Decoder} to achieve a controllable deep image prior. Network parameters are initialized with He initialization \citep{he2015delving}. 
Our approach works with popular optimizers such as standard gradient descent and Adam \citep{kingma2014adam}. Following \cite{ulyanov2018deep,ulyanov2020deep}, we use Adam with a mini-batch of 1 to optimize our networks.
We set $\beta_1$ to 0.9, $\beta_2$ to 0.999 and the initial learning rate to 0.001. The network input is a uniform noise between 0 and 0.1 with a depth of 32 by default. Our code will be released.

\subsection{Image Denoising}
\label{sec_denoising}
For the denoising comparison we use two datasets, \ie, the standard dataset by~\cite{dabov2007image} consisting of 9 RGB images, and CBSD68 by~\cite{roth2009fields} consisting of 68 RGB images. Each noisy image is generated by adding an additive Gaussian white noise with three noise levels, including $\sigma{=}15$, $\sigma{=}25$ and $\sigma{=}50$. The goal is to distill the original image without Gaussian noise. Results on the dataset of \cite{dabov2007image} are shown in Fig. \ref{fig:exp-2-1}, where PSNR scores of various methods are shown over multiple iterations. The performance of the deep image prior~\citep{ulyanov2018deep,ulyanov2020deep} gradually degrades after reaching a peak. For each image, the peak is reached at a different number of iterations, so simply using a fixed number of iterations will be suboptimal for most images.

\begin{table}[!h]
\small
\caption{\textbf{Image denoising} on CBSD68 for varying $\sigma$. Supervised approaches and CBM3D prevail, but our unsupervised method obtains better PSNR than the deep image prior and its variants across three noise levels.}
\centering
\resizebox{1\linewidth}{!}{
\begin{threeparttable}
\setlength\tabcolsep{2.5pt}
\begin{tabular}{lccc}
\toprule
& 15 & 25 & 50 \\
\hline
\cite{ulyanov2020deep}$^{\dagger *}$& 30.58 & 27.84 & 24.59  \\
\cite{heckel2018deep}$^{\dagger}$ & 28.66 & 26.60 & 24.06 \\
\cite{cheng2019bayesian}$^{\dagger}$ & 30.77 & 28.08 & 24.71\\
\textit{Ours} & \textbf{30.80} & \textbf{28.15} & \textbf{24.83}\\
\hline
CBM3D \citep{dabov2007image}$^{\dagger \dagger}$ & 33.52 & 30.71 & 27.38\\
CDnCNN \citep{zhang2017beyond}$^{\dagger \dagger}$ & 33.89 & 31.23 & 27.92\\
FFDNet \citep{zhang2018ffdnet} & 33.87 & 31.21 & 27.96\\
\bottomrule
\end{tabular}
\footnotesize{$^{\dagger}$Results based on author-provided code.\\ $^*$Results obtained with oracle stopping. \\$^{\dagger \dagger}$Results provided by \cite{zhang2018ffdnet}.}
  \end{threeparttable}
}
\label{tab_noise}
\end{table}
Our method provides two advantages: 1) The performance does not decay over iterations with controlled spectral bias; 2) The optimization can be automatically stopped at an appropriate moment using the proposed stopping criterion, leading to good PSNR scores for all images (marked by the green vertical lines). \cite{heckel2018deep} achieve fast convergence without performance degradation, but at the expense of reduced performance. \cite{cheng2019bayesian} obtain comparable PSNR scores, but they require 2 to 4 times as many iterations to converge. 

So far, we have shown the performance of various methods per image over a varying number of optimization iterations. Next, we compare their overall PSNR performance on the 68 images in CBSD68, as shown in Table \ref{tab_noise}. 
While our unsupervised method is outperformed by supervised-learning alternatives \citep{zhang2017beyond,zhang2018ffdnet} and CBM3D \citep{dabov2007image}, it does better than the deep image prior \citep{ulyanov2018deep,ulyanov2020deep}, and its variants \citep{heckel2018deep,cheng2019bayesian} across three noise levels. 
We also provide qualitative results for denoising in Fig.~\ref{exp_fig_denoise}, where we observe our method preserves the high-frequency edges without overfitting to high-frequency noise.
\begin{table}[t!]
\caption{\textbf{JPEG image deblocking} on LIVE1 for varying quality levels. Supervised approached prevail, but compared to the unsupervised deep image prior and its variants, our method obtains better performance in terms of PSNR.}
\centering
\resizebox{1\linewidth}{!}{
\begin{threeparttable}
\setlength\tabcolsep{2.5pt}
\begin{tabular}{lccc}
\toprule
& 10 & 20 & 30\\
\hline
\cite{ulyanov2020deep}$^{\dagger *}$& 27.52 & 29.75 & 31.08\\
\cite{heckel2018deep}$^{\dagger}$ & 26.63 & 27.65 & 28.99 \\
\cite{cheng2019bayesian}$^{\dagger}$ & 27.59 & 29.81 & 31.12\\
\textit{Ours} & \textbf{27.70} & \textbf{29.86} & \textbf{31.14}\\
\hline
AR-CNN \citep{dong2015compression}$^{\dagger \dagger}$ & 28.96 & 31.29 & 32.67\\
TNRD \citep{chen2016trainable}$^{\dagger \dagger}$ & 29.15 & 31.46 & 32.84\\
DnCNN \citep{zhang2017beyond} & 29.19 & 31.59 & 32.98\\
\bottomrule
\end{tabular}
\footnotesize{$^{\dagger}$Results based on author-provided code. \\ $^*$Results obtained with oracle stopping. \\ $^{\dagger \dagger}$Results provided by \cite{zhang2017beyond}.}
  \end{threeparttable}
}
\label{tab_deblocking}
\end{table}
\begin{table}[h!]
\small
\caption{\textbf{Image inpainting} on CBSD68 for varying ratio. In terms of PSNR, our method outperforms the deep image prior and its variants on region-based inpainting, across three hole-to-image area ratios.}
\centering
\resizebox{1\linewidth}{!}{
\begin{threeparttable}
\setlength\tabcolsep{2.5pt}
\begin{tabular}{lccc}
\toprule
& 0.1 & 0.25 & 0.5 \\

\hline
\cite{ulyanov2020deep} $^{\dagger *}$& 22.78 & 19.42 & 17.26  \\
\cite{heckel2018deep}$^{\dagger}$ & 21.52 & 18.67 & 16.81 \\
\cite{cheng2019bayesian}$^{\dagger}$ & 22.83 & 19.49 & 17.28\\
\textit{Ours} & \textbf{22.87} & \textbf{19.58} & \textbf{17.36}\\
\bottomrule
\end{tabular}
\footnotesize{$^{\dagger}$Results based on author-provided code.\\ $^*$Results obtained with oracle stopping.} 
  \end{threeparttable}
}
\label{tab_inpainting}
\end{table}

\subsection{JPEG Image Deblocking}
JPEG image deblocking is the process of reducing the compression artifacts in JPEG images. We evaluate on the Classic5 dataset by \cite{foi2006pointwise} and the LIVE1 dataset by \cite{sheikh2006statistical}. Classic5  consists of $5$ gray-scale images, and LIVE1 consists of 29 color images. Following \cite{dong2015compression}, the color images are transformed to gray-scale using the YCbCr color model by keeping the Y component only. Then, the gray-scale images are compressed with the PIL encoder into three qualities, $10$, $20$, and $30$. Fig. \ref{fig:exp-2-2} provides a quantitative comparison on Classic5 for  $quality{=}10$. Akin to the denoising comparison, we again observe the degradation of performance over iterations for the deep image prior \citep{ulyanov2018deep,ulyanov2020deep}. \cite{cheng2019bayesian} and \cite{heckel2018deep} do not suffer from the degradation problem, at the expense of either reduced performance or slow convergence. With the controlled spectral bias and automatic stopping criterion, we achieve a good trade-off between PSNR score and convergence (marked by the green vertical lines).

We also provide quantitative results for LIVE1 in Table \ref{tab_deblocking}. 
Naturally, the learning-based methods \citep{dong2015compression,chen2016trainable,zhang2017beyond} perform best. Across three quality levels, our unsupervised method performs better than the deep image prior \citep{ulyanov2018deep,ulyanov2020deep} and its two variants \citep{heckel2018deep,cheng2019bayesian}. We also provide qualitative examples in Fig.~\ref{exp_fig_deblocking}, which shows that our method better reduces the artifacts and recovers high-frequency image details.

\subsection{Image Inpainting}
In image inpainting, we are given an image with missing pixels resulting from a binary mask. The goal is to reconstruct the missing data. We evaluate on the standard dataset by \cite{heide2015fast}, consisting of 11 grayscale images, and the CBSD68 dataset by~\citep{roth2009fields} consisting of 68 RGB images. Following \cite{ulyanov2018deep,ulyanov2020deep,cheng2019bayesian}, we consider inpainting with masks that are randomly sampled according to a binary Bernoulli distribution on the standard dataset. Each mask is sampled to drop 50\% of the pixels at random. For CBSD68, we consider inpainting with central region masks and we evaluate on three hole-to-image area ratios, $ratio{=}0.1$, $ratio{=}0.25$ and $ratio{=}0.5$, following \cite{pathak2016context}. Fig. \ref{fig:exp-2-3} provides a quantitative comparison on the standard dataset. We also provide quantitative results for CBSD68 in Table \ref{tab_inpainting}. Our observations are the same as for the denoising and deblocking comparison.
We provide qualitative examples for pixel inpainting in Fig.~\ref{exp_fig_inpainting} and region inpainting in Fig.~\ref{exp_fig_hole_inpainting}, which shows our ability to recover high-frequency details.

\subsection{Super-resolution}
In image super-resolution, a low-resolution image is given; the goal is to recover its scaled-up version. Following \cite{ulyanov2018deep,ulyanov2020deep}, the network generates a high-resolution image from the random noise input. The high-resolution image is then downsampled using a differentiable Lanczos filter to compute the loss with the provided low resolution image for optimizing the network. We report on the standard Set14 dataset by \cite{zeyde2010single} and Set5 by \cite{bevilacqua2012low}. We evaluate the performance for an up-scaling of 4 and 8.
For the super-resolution task, the deep image prior \citep{ulyanov2018deep,ulyanov2020deep} does not suffer from the performance degradation over iterations because the optimization objective strives to find the low-resolution image without high-frequency noise.
Following \cite{ulyanov2018deep,ulyanov2020deep}, we report the PSNR score at a stopping iteration of 2,000 for the scaling of 4, and 4,000 for the scaling of 8. Results on Set 14 are provided in Table \ref{tab_sr8} and results on Set 5 are summarized in Table \ref{tab_sr4}. On most images our method achieves better performance, not only for \cite{ulyanov2018deep,ulyanov2020deep} but also
compared to \cite{heckel2018deep} and \cite{cheng2019bayesian}.
We provide a qualitative comparison in Fig.~\ref{exp_fig_sr}. We observe that our method produces fewer high-frequency artifacts than \cite{ulyanov2018deep,ulyanov2020deep} and \cite{cheng2019bayesian}. We postulate that our Lipschitz normalization contributes to the benefit. Interestingly, our method also recovers fine details. A likely explanation is that our Gaussian upsampling is better at learning the desired higher frequencies. Note that fine details like textures are high-frequency compared to flat regions, but still relatively low-frequency compared to most artifacts.

\begin{table*}[h!]
\small
\caption{\textbf{Super-resolution} on Set14. The PSNR scores are reported for a stopping iteration of 2,000 for the scaling of 4, and 4,000 for the scaling of 8, following \cite{ulyanov2018deep,ulyanov2020deep}. On most images we achieve better performance than existing methods, and we obtain the highest PSNR on average for $4\times$ and $8\times$ super-resolution. }
\centering
\resizebox{1\linewidth}{!}{
\begin{threeparttable}
\setlength\tabcolsep{2.5pt}
\begin{tabular}{@{}lccccccccccccccc@{}}
\toprule
 & Baboon & Barbara & Bridge & Coastguard & Comic & Face & Flowers & Foreman & Lenna& Man & Monarch & Pepper & Ppt3 & Zebra & \textit{Average} \\
\hline
\rowcolor{Gray}
\textbf{$4 \times$ resolution} & & & & & & & & & & & & & & & \\
\cite{ulyanov2020deep}$^*$& 22.29 &25.53 &24.38 & 25.81 & 22.18 & 31.02 & 26.14 & 31.66 & 30.83 & 26.09 & 29.98 & 32.08 & 24.38&25.71 &27.00 \\
\cite{heckel2018deep}$^{\dagger}$ & 20.54 &21.51&20.97 &23.52 & 18.86 & 28.15 & 20.88 & 24.44 & 24.07 & 21.18 & 21.21 &23.89 & 17.28 & 18.59 & 21.79 \\
\cite{cheng2019bayesian}$^{\dagger}$ & 21.51 &24.84 & 23.74 & 25.02 & 21.94 & 30.11 & 25.41 & 30.57 & 28.62 &25.37 & 28.41 & 30.25 & 23.69 &24.48 & 26.07 \\
\textit{Ours} & \textbf{22.81} & \textbf{25.74}& \textbf{25.02} & \textbf{25.86} & \textbf{22.31} & \textbf{32.09} & \textbf{26.74} & \textbf{32.77} & \textbf{31.29} &\textbf{26.42} & \textbf{30.77} & \textbf{32.62} & \textbf{24.73} &\textbf{25.87} & \textbf{27.50} \\
\hline
Bicubic$^{\dagger \dagger}$ &22.44 &25.15 &24.47 &25.53 &21.59 &31.34 &25.33 &29.45 &29.84 &25.70 &27.45 &30.63 &21.78 &24.01 &26.05\\
TV prior$^{\dagger \dagger}$ &22.34 &24.78 &24.46 &25.78 &21.95 &31.34 &25.91 &30.63 &29.76 &25.94 &28.46 &31.32 &22.75 &24.52 &26.42 \\
LapSRN \citep{lai2018fast}$^{\dagger \dagger}$ &22.83 &25.69 &25.36 &26.21 &22.90 &32.62 &27.54 &33.59 &31.98 &27.27 &31.62 &33.88 &25.36 &26.98 &28.13\\
\midrule
\rowcolor{Gray}
\textbf{$8 \times$ resolution} & & & & & & & & & & & & & & & \\
\cite{ulyanov2020deep}$^*$& 21.38 &23.94 & 22.20 & \textbf{24.21} & 19.86 & 29.52 & 22.86 & 27.87 & \textbf{27.93} & 23.57 &\textbf{24.86}  & \textbf{29.18} & 20.12 &\textbf{20.62} &24.15 \\
\cite{heckel2018deep}$^{\dagger}$ & 20.07 &19.86 & 19.67 & 22.30 &18.01 & 26.53 & 19.57 & 22.76 & 22.06 & 19.54& 19.71 & 21.44 & 15.64 & 17.20&20.31 \\
\cite{cheng2019bayesian}$^{\dagger}$& 19.81 &23.69 & 22.19 & 19.22 & 19.72& 28.88 & 22.81 & 27.34 & 19.69 & 23.36&24.43 & 28.72 &26.14  & 19.89 &20.67\\
 \textit{Ours} & \textbf{21.57} &\textbf{24.48} & \textbf{22.64} & 24.18 & \textbf{19.71} & \textbf{29.94} & \textbf{22.92} & \textbf{27.95} & 27.67 &\textbf{23.86} & 24.46 & 28.91 & \textbf{23.28} & 19.93 & \textbf{24.17} \\
 \hline
Bicubic$^{\dagger \dagger}$ &21.28 &23.44 &22.24 &23.65 &19.25 &28.79 &22.06 &25.37 &26.27 &23.06 &23.18 &26.55 &18.62 &19.59 &23.09\\
TV prior$^{\dagger \dagger}$ &21.30 &23.72 &22.30 &23.82 &19.50 &28.84 &22.50 &26.07 &26.74 &23.53 &23.71 &27.56 &19.34 &19.89 &23.48\\
LapSRN \citep{lai2018fast}$^{\dagger \dagger}$ &21.51 &24.21 &22.77 &24.10 &20.06 &29.85 &23.31 &28.13 &28.22 &24.20 &24.97 &29.22 &20.13 &20.28 &24.35 \\
\bottomrule
\end{tabular}
\footnotesize{$^{\dagger}$Results based on author-provided code. $^*$Results obtained with oracle stopping. $^{\dagger \dagger}$Results provided by \cite{ulyanov2020deep}.}
  \end{threeparttable}
}
\label{tab_sr8}
\end{table*}
\begin{table}[h!]
\small
\caption{\textbf{Super-resolution} on set5. The PSNR scores are reported for a stopping iteration of 2,000 for the scaling of 4, and 4,000 for the scaling of 8, following \cite{ulyanov2020deep}. On most images we achieve better performance than existing methods, and we perform best on average for both $4\times$ and $8\times$ super-resolution.}
\centering
\resizebox{1\linewidth}{!}{
\begin{threeparttable}
\setlength\tabcolsep{2.5pt}
\begin{tabular}{lcccccccc}
\toprule
& Baby & Bird & Butterfly & Head & Woman & \textit{Average} \\
\rowcolor{Gray}
\textbf{$4 \times$ resolution} & & & & & & \\
\cite{ulyanov2020deep} $^*$& 31.49 & 31.80 & 26.23 & 31.04 & \textbf{28.93} &29.89 \\
\cite{heckel2018deep}$^{\dagger}$ & 24.57 & 24.66 & 18.46 & 27.64 & 22.44  &23.55 \\
\cite{cheng2019bayesian}$^{\dagger}$ & 27.35 & 28.37 & 24.21 & 27.45 & 25.48 &26.57\\
\textit{Ours}& \textbf{32.76} & \textbf{32.71} & \textbf{26.47} & \textbf{31.79} & 28.54 &\textbf{30.45} \\
\hline
Bicubic $^{\dagger \dagger}$ &31.78 &30.20 &22.13 &31.34 &26.75 &28.44\\
TV prior $^{\dagger \dagger}$ &31.21 &30.43 &24.38 &31.34 &26.93 &28.85\\
LapSRN \citep{lai2018fast} $^{\dagger \dagger}$ &33.55 &33.76 &27.28 &32.62 &30.72 &31.58 \\
\midrule
\rowcolor{Gray}
\textbf{$8 \times$ resolution} & & & & & & \\
\cite{ulyanov2020deep} $^*$& 28.28 & \textbf{27.09} & 20.02 &29.55 & 24.50 &25.88 \\
\cite{heckel2018deep}$^{\dagger}$ & 21.95 & 22.97 &16.18 & 26.56 & 20.46 &21.62 \\
\cite{cheng2019bayesian}$^{\dagger}$ & 27.18 & 26.64 &19.64 & 24.76 & 23.81 &24.41\\
\textit{Ours}& \textbf{28.46} & 26.79 &\textbf{20.32} & \textbf{30.07} &\textbf{24.76} &\textbf{26.08} \\
\hline
Bicubic $^{\dagger \dagger}$ &27.28 &25.28 &17.74 &28.82 &22.74 &24.37\\
TV prior $^{\dagger \dagger}$ &27.93 &25.82 &18.40 &28.87 &23.36 &24.87\\
LapSRN \citep{lai2018fast} $^{\dagger \dagger}$ &28.88 &27.10 &19.97 &29.76 &24.79 &26.10\\
\bottomrule
\end{tabular}
\footnotesize{$^{\dagger}$Results based on author-provided code.\\ $^*$Results obtained with oracle stopping. \\$^{\dagger \dagger}$Results provided by \cite{ulyanov2020deep}.}
  \end{threeparttable}
}
\label{tab_sr4}
\end{table}

\subsection{Image Enhancement}
Following \cite{ulyanov2018deep,ulyanov2020deep}, we also evaluate our method on image enhancement. The deep image prior performs sharpness enhancement by means of unsharp masking \citep{morishita1988unsharp,shi2021unsharp}, which can be described by $x_e = (x_0-x_s) + x_0$, where an enhanced image is represented by $x_e$, an original image by $x_0$, an unsharp mask by $(x_0-x_s)$ where $x_s$ denotes the smoothed version of the original image. The smoothness of $x_s$ controls the size of the region around the edge pixels that is affected by sharpening. The higher the smoothness, the wider the regions around the edges that got sharpened.
The deep image prior obtains the smoothed images by stopping the optimization at different iterations. However, the smoothness of the output image is quite sensitive to the number of optimization iterations, which is hard to control. By contrast, our method is able to manipulate the smoothness of the output image by tuning $\lambda$ in Eq. (\ref{eq_lipnorm}). 
Thus, we obtain the smoothed images with different $\lambda$, by optimizing the network in a fixed iteration of $5,000$.
The smaller the $\lambda$, the higher the smoothness of the output images and the more enhancement to the image details, as shown in Fig. \ref{exp_fig_enhancement}.

\subsection{Success and Failure Cases}
\begin{figure}[t!]
\centering
\begin{subfigure}{0.235\textwidth}
\includegraphics[width=\textwidth]{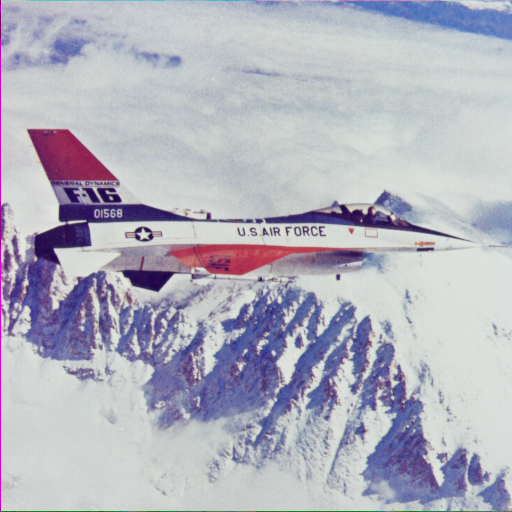}
\caption{\textbf{Clean image}}
\label{exp_fig_fail-a} 
\end{subfigure}
\begin{subfigure}{0.235\textwidth}
\includegraphics[width=\textwidth]{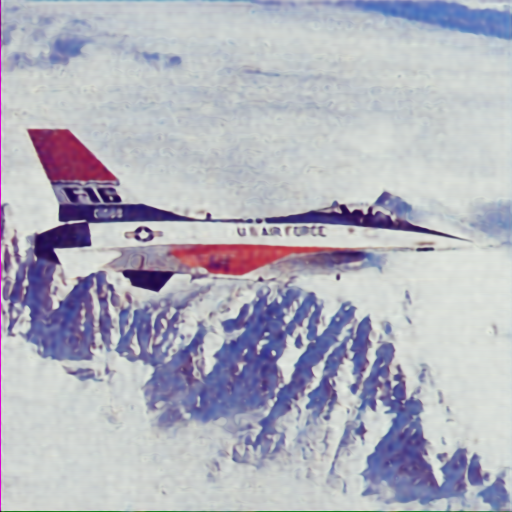}
\caption{\textbf{$\sigma=25$}}
\label{exp_fig_fail-b} 
\end{subfigure}
\begin{subfigure}{0.235\textwidth}
\includegraphics[width=\textwidth]{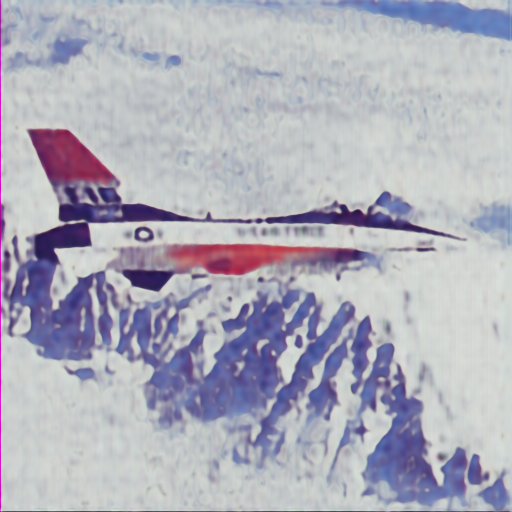}
\caption{\textbf{$\sigma=50$}}
\label{exp_fig_fail-c} 
\end{subfigure}
\begin{subfigure}{0.235\textwidth}
\includegraphics[width=\textwidth]{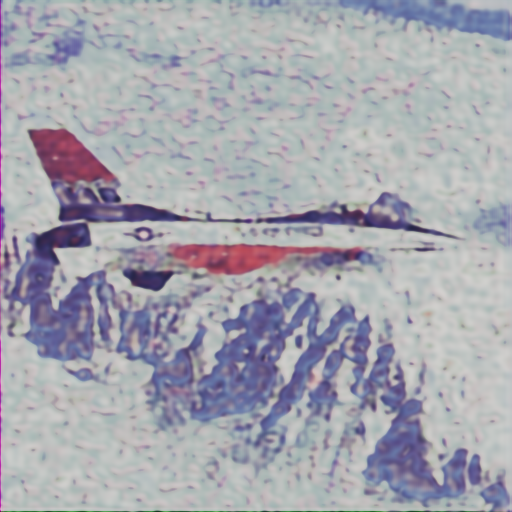}
\caption{\textbf{$\sigma=100$}}
\label{exp_fig_fail-d} 
\end{subfigure}
\caption{\textbf{Success and failure case} of our method for image denoising on image `F16'. Our method performs well when the noise level is modest ($\sigma{=}25$), while it fails to remove noise when the noise level is too high 
($\sigma{=}100$).}
\label{exp_fig_fail}    
\end{figure}

We return to the denoising task to analyze a success and failure case or our approach in Fig.~\ref{exp_fig_fail}. The goal is to remove additive Gaussian noise from a natural image. Our method performs well when the noise level is modest, as shown in Fig.~\ref{exp_fig_fail-b}. However, with higher noise levels, the proposed method fails to remove the noise, as shown in Fig.~\ref{exp_fig_fail-d}. We attribute this to the fact that in the frequency domain, additive Gaussian noise has equal intensity at different frequencies. By contrast, the power spectrum of a natural image decays rapidly from low frequencies to high frequencies \citep{ruderman1994statistics}. Consequently, when the noise level is low, noise is usually dominant at high frequencies and the natural signal is more dominant at lower frequencies. However, the noise can also be more dominant at lower frequencies with higher level. In this case, separating low-frequencies from high-frequencies through spectral bias fails to remove the noise. 

%% file: 6-conclusion.tex
\section{Conclusion}
In this paper, we show the spectral bias leads inverse imaging networks to capture the deep image prior during optimization, independent of their architectures. We do so by introducing a metric, the Frequency Band Correspondence, which offers a spectral measurement of the deep image prior, revealing the low frequency natural image signals are learned faster and better than high-frequency noise signals. We also introduce Lipschitz normalization and Gaussian upsampling that allow to manipulate and adjust the spectral bias for inverse imaging problems. Besides these methods for controlling spectral bias, we further introduce a simple automatic stopping criterion to avoid superfluous computation.
The experiments show that our method does not suffer from the performance degradation over iterations with controlled spectral bias and enables stopping the optimization automatically at an appropriate moment using the proposed stopping criterion. Our method also obtains favorable performance compared to current approaches for denoising, deblocking, inpainting, super-resolution and detail enhancement.

%% file: 0-main.bbl
\begin{thebibliography}{73}
\providecommand{\natexlab}[1]{#1}
\providecommand{\url}[1]{{#1}}
\providecommand{\urlprefix}{URL }
\expandafter\ifx\csname urlstyle\endcsname\relax
  \providecommand{\doi}[1]{DOI~\discretionary{}{}{}#1}\else
  \providecommand{\doi}{DOI~\discretionary{}{}{}\begingroup
  \urlstyle{rm}\Url}\fi
\providecommand{\eprint}[2][]{\url{#2}}

\bibitem[{Arias et~al.(2011)Arias, Facciolo, Caselles, and
  Sapiro}]{arias2011variational}
Arias P, Facciolo G, Caselles V, Sapiro G (2011) A variational framework for
  exemplar-based image inpainting. International Journal of Computer Vision
  93(3):319--347

\bibitem[{Arridge et~al.(2019)Arridge, Maass, {\"O}ktem, and
  Sch{\"o}nlieb}]{arridge2019solving}
Arridge S, Maass P, {\"O}ktem O, Sch{\"o}nlieb CB (2019) Solving inverse
  problems using data-driven models. Acta Numerica 28:1--174

\bibitem[{Asim et~al.(2019)Asim, Shamshad, and Ahmed}]{asim2019patchdip}
Asim M, Shamshad F, Ahmed A (2019) Patchdip exploiting patch redundancy in deep
  image prior for denoising. In: NeurIPS Workshop on Solving Inverse Problems
  with Deep Networks

\bibitem[{Bahrami and Kot(2014)}]{bahrami2014fast}
Bahrami K, Kot AC (2014) A fast approach for no-reference image sharpness
  assessment based on maximum local variation. IEEE Signal Processing Letters
  21(6):751--755

\bibitem[{Bertero and Boccacci(1998)}]{bertero1998introduction}
Bertero M, Boccacci P (1998) Introduction to inverse problems in imaging. IOP
  Publishing

\bibitem[{Bevilacqua et~al.(2012)Bevilacqua, Roumy, Guillemot, and
  Alberi-Morel}]{bevilacqua2012low}
Bevilacqua M, Roumy A, Guillemot C, Alberi-Morel ML (2012) Low-complexity
  single-image super-resolution based on nonnegative neighbor embedding. In:
  BMVC

\bibitem[{Chakrabarty and Maji(2019)}]{chakrabarty2019spectral}
Chakrabarty P, Maji S (2019) The spectral bias of the deep image prior. In:
  NeurIPS Workshop on Bayesian Deep Learning

\bibitem[{Chen et~al.(2020{\natexlab{a}})Chen, Fan, Liao, Aviles-Rivero, Yuan,
  Yu, and Hua}]{chen2020controllable}
Chen D, Fan Q, Liao J, Aviles-Rivero A, Yuan L, Yu N, Hua G
  (2020{\natexlab{a}}) Controllable image processing via adaptive filterbank
  pyramid. IEEE Transactions on Image Processing 29:8043--8054

\bibitem[{Chen and Pock(2016)}]{chen2016trainable}
Chen Y, Pock T (2016) Trainable nonlinear reaction diffusion: A flexible
  framework for fast and effective image restoration. IEEE Transactions on
  Pattern Analysis and Machine Intelligence 39(6):1256--1272

\bibitem[{Chen et~al.(2020{\natexlab{b}})Chen, Gao, Robb, and
  Huang}]{chen2020dip}
Chen YC, Gao C, Robb E, Huang JB (2020{\natexlab{b}}) Nas-dip: Learning deep
  image prior with neural architecture search. In: ECCV

\bibitem[{Cheng et~al.(2019)Cheng, Gadelha, Maji, and
  Sheldon}]{cheng2019bayesian}
Cheng Z, Gadelha M, Maji S, Sheldon D (2019) A bayesian perspective on the deep
  image prior. In: CVPR

\bibitem[{Crete et~al.(2007)Crete, Dolmiere, Ladret, and
  Nicolas}]{crete2007blur}
Crete F, Dolmiere T, Ladret P, Nicolas M (2007) The blur effect: Perception and
  estimation with a new no-reference perceptual blur metric. In: SPIE HVEI

\bibitem[{Dabov et~al.(2007)Dabov, Foi, Katkovnik, and
  Egiazarian}]{dabov2007image}
Dabov K, Foi A, Katkovnik V, Egiazarian K (2007) Image denoising by sparse 3-d
  transform-domain collaborative filtering. IEEE Transactions on Image
  Processing 16(8):2080--2095

\bibitem[{Dai et~al.(2020)Dai, Feng, Wu, Chen, Lu, Jiang, and
  Xia}]{dai2020dipdefend}
Dai T, Feng Y, Wu D, Chen B, Lu J, Jiang Y, Xia ST (2020) {DIPDefend}: Deep
  image prior driven defense against adversarial examples. In: ACM MM

\bibitem[{Daubechies et~al.(2004)Daubechies, Defrise, and
  De~Mol}]{daubechies2004iterative}
Daubechies I, Defrise M, De~Mol C (2004) An iterative thresholding algorithm
  for linear inverse problems with a sparsity constraint. Communications on
  Pure and Applied Mathematics 57(11):1413--1457

\bibitem[{Dong et~al.(2015{\natexlab{a}})Dong, Deng, Loy, and
  Tang}]{dong2015compression}
Dong C, Deng Y, Loy CC, Tang X (2015{\natexlab{a}}) Compression artifacts
  reduction by a deep convolutional network. In: ICCV, pp 576--584

\bibitem[{Dong et~al.(2015{\natexlab{b}})Dong, Loy, He, and
  Tang}]{dong2015image_pami}
Dong C, Loy CC, He K, Tang X (2015{\natexlab{b}}) Image super-resolution using
  deep convolutional networks. IEEE Transactions on Pattern Analysis and
  Machine Intelligence 38(2):295--307

\bibitem[{Dong et~al.(2015{\natexlab{c}})Dong, Shi, Ma, and Li}]{dong2015image}
Dong W, Shi G, Ma Y, Li X (2015{\natexlab{c}}) Image restoration via
  simultaneous sparse coding: Where structured sparsity meets gaussian scale
  mixture. International Journal of Computer Vision 114(2):217--232

\bibitem[{Efros and Leung(1999)}]{efros1999texture}
Efros AA, Leung TK (1999) Texture synthesis by non-parametric sampling. In:
  ICCV, vol~2, pp 1033--1038

\bibitem[{Elad et~al.(2010)Elad, Figueiredo, and Ma}]{elad2010role}
Elad M, Figueiredo MA, Ma Y (2010) On the role of sparse and redundant
  representations in image processing. Proceedings of the IEEE 98(6):972--982

\bibitem[{Engl et~al.(1996)Engl, Hanke, and Neubauer}]{engl1996regularization}
Engl HW, Hanke M, Neubauer A (1996) Regularization of inverse problems, vol
  375. Springer Science \& Business Media

\bibitem[{Foi et~al.(2006)Foi, Katkovnik, and Egiazarian}]{foi2006pointwise}
Foi A, Katkovnik V, Egiazarian K (2006) Pointwise shape-adaptive dct for
  high-quality deblocking of compressed color images. In: ESPC, pp 1--5

\bibitem[{Gandelsman et~al.(2019)Gandelsman, Shocher, and
  Irani}]{gandelsman2019double}
Gandelsman Y, Shocher A, Irani M (2019) " double-dip": Unsupervised image
  decomposition via coupled deep-image-priors. In: CVPR

\bibitem[{Hahn et~al.(2011)Hahn, Tai, Borok, and
  Bruckstein}]{hahn2011orientation}
Hahn J, Tai XC, Borok S, Bruckstein AM (2011) Orientation-matching minimization
  for image denoising and inpainting. International Journal of Computer Vision
  92(3):308--324

\bibitem[{He et~al.(2015)He, Zhang, Ren, and Sun}]{he2015delving}
He K, Zhang X, Ren S, Sun J (2015) Delving deep into rectifiers: Surpassing
  human-level performance on imagenet classification. In: ICCV

\bibitem[{Heckel and Hand(2019)}]{heckel2018deep}
Heckel R, Hand P (2019) Deep decoder: Concise image representations from
  untrained non-convolutional networks. In: ICLR

\bibitem[{Heckel and Soltanolkotabi(2020)}]{heckel2019denoising}
Heckel R, Soltanolkotabi M (2020) Denoising and regularization via exploiting
  the structural bias of convolutional generators. In: ICLR

\bibitem[{Heide et~al.(2015)Heide, Heidrich, and Wetzstein}]{heide2015fast}
Heide F, Heidrich W, Wetzstein G (2015) Fast and flexible convolutional sparse
  coding. In: CVPR

\bibitem[{Ho et~al.(2020)Ho, Gilbert, Jin, and Collomosse}]{ho2020neural}
Ho K, Gilbert A, Jin H, Collomosse J (2020) Neural architecture search for deep
  image prior. ArXiv:200104776

\bibitem[{Ioffe and Szegedy(2015)}]{ioffe2015batch}
Ioffe S, Szegedy C (2015) Batch normalization: Accelerating deep network
  training by reducing internal covariate shift. In: ICML

\bibitem[{Jain and Seung(2008)}]{jain2008natural}
Jain V, Seung S (2008) Natural image denoising with convolutional networks

\bibitem[{Jin et~al.(2017)Jin, McCann, Froustey, and Unser}]{jin2017deep}
Jin KH, McCann MT, Froustey E, Unser M (2017) Deep convolutional neural network
  for inverse problems in imaging. IEEE Transactions on Image Processing
  26(9):4509--4522

\bibitem[{Katsaggelos(1989)}]{katsaggelos1989iterative}
Katsaggelos AK (1989) Iterative image restoration algorithms. Optical
  Engineering 28(7):287735

\bibitem[{Kattamis et~al.(2019)Kattamis, Adel, and
  Weller}]{kattamis2019exploring}
Kattamis A, Adel T, Weller A (2019) Exploring properties of the deep image
  prior. In: NeurIPS Workshop on Solving Inverse Problems with Deep Networks

\bibitem[{Katznelson(2004)}]{katznelson2004introduction}
Katznelson Y (2004) An introduction to harmonic analysis. Cambridge University
  Press

\bibitem[{Kindermann et~al.(2005)Kindermann, Osher, and
  Jones}]{kindermann2005deblurring}
Kindermann S, Osher S, Jones PW (2005) Deblurring and denoising of images by
  nonlocal functionals. Multiscale Modeling \& Simulation 4(4):1091--1115

\bibitem[{Kingma and Ba(2015)}]{kingma2014adam}
Kingma DP, Ba J (2015) Adam: A method for stochastic optimization. In: ICLR

\bibitem[{Lai et~al.(2018)Lai, Huang, Ahuja, and Yang}]{lai2018fast}
Lai WS, Huang JB, Ahuja N, Yang MH (2018) Fast and accurate image
  super-resolution with deep laplacian pyramid networks. IEEE transactions on
  pattern analysis and machine intelligence 41(11):2599--2613

\bibitem[{Ledig et~al.(2017)Ledig, Theis, Husz{\'a}r, Caballero, Cunningham,
  Acosta, Aitken, Tejani, Totz, Wang, and Shi}]{ledig2017photo}
Ledig C, Theis L, Husz{\'a}r F, Caballero J, Cunningham A, Acosta A, Aitken A,
  Tejani A, Totz J, Wang Z, Shi W (2017) Photo-realistic single image
  super-resolution using a generative adversarial network. In: CVPR, pp
  4681--4690

\bibitem[{Lefkimmiatis(2018)}]{lefkimmiatis2018universal}
Lefkimmiatis S (2018) Universal denoising networks: a novel cnn architecture
  for image denoising. In: CVPR, pp 3204--3213

\bibitem[{Li et~al.(2018)Li, You, and Robles-Kelly}]{li2018frequency}
Li J, You S, Robles-Kelly A (2018) A frequency domain neural network for fast
  image super-resolution. In: IJCNN, pp 1--8

\bibitem[{Lin et~al.(2008)Lin, He, Tang, and Tang}]{lin2008limits}
Lin Z, He J, Tang X, Tang CK (2008) Limits of learning-based superresolution
  algorithms. International Journal of Computer Vision 80(3):406--420

\bibitem[{Liu et~al.(2019)Liu, Sun, Xu, and Kamilov}]{liu2019image}
Liu J, Sun Y, Xu X, Kamilov US (2019) Image restoration using total variation
  regularized deep image prior. In: ICASSP

\bibitem[{Lucas et~al.(2018)Lucas, Iliadis, Molina, and
  Katsaggelos}]{lucas2018using}
Lucas A, Iliadis M, Molina R, Katsaggelos AK (2018) Using deep neural networks
  for inverse problems in imaging: beyond analytical methods. IEEE Signal
  Processing Magazine 35(1):20--36

\bibitem[{Mairal et~al.(2009)Mairal, Bach, Ponce, Sapiro, and
  Zisserman}]{mairal2009non}
Mairal J, Bach F, Ponce J, Sapiro G, Zisserman A (2009) Non-local sparse models
  for image restoration. In: ICCV, pp 2272--2279

\bibitem[{Mao et~al.(2016)Mao, Shen, and Yang}]{mao2016image}
Mao X, Shen C, Yang YB (2016) Image restoration using very deep convolutional
  encoder-decoder networks with symmetric skip connections. In: NeurIPS

\bibitem[{Mataev et~al.(2019)Mataev, Milanfar, and Elad}]{mataev2019deepred}
Mataev G, Milanfar P, Elad M (2019) Deepred: Deep image prior powered by red.
  In: ICCV Workshop on Learning for Computational Imaging

\bibitem[{McCann et~al.(2017)McCann, Jin, and Unser}]{mccann2017convolutional}
McCann MT, Jin KH, Unser M (2017) Convolutional neural networks for inverse
  problems in imaging: A review. IEEE Signal Processing Magazine 34(6):85--95

\bibitem[{Miyato et~al.(2018)Miyato, Kataoka, Koyama, and
  Yoshida}]{miyato2018spectral}
Miyato T, Kataoka T, Koyama M, Yoshida Y (2018) Spectral normalization for
  generative adversarial networks. In: ICLR

\bibitem[{Morishita et~al.(1988)Morishita, Yamagata, Okabe, Yokoyama, and
  Hamatani}]{morishita1988unsharp}
Morishita K, Yamagata S, Okabe T, Yokoyama T, Hamatani K (1988) Unsharp masking
  for image enhancement. US Patent 4,794,531

\bibitem[{Olshausen and Field(1996)}]{olshausen1996emergence}
Olshausen BA, Field DJ (1996) Emergence of simple-cell receptive field
  properties by learning a sparse code for natural images. Nature
  381(6583):607--609

\bibitem[{Pathak et~al.(2016)Pathak, Krahenbuhl, Donahue, Darrell, and
  Efros}]{pathak2016context}
Pathak D, Krahenbuhl P, Donahue J, Darrell T, Efros AA (2016) Context encoders:
  feature learning by inpainting. In: CVPR, pp 2536--2544

\bibitem[{Portilla(2009)}]{portilla2009image}
Portilla J (2009) Image restoration through l0 analysis-based sparse
  optimization in tight frames. In: ICIP, pp 3909--3912

\bibitem[{Protter et~al.(2008)Protter, Elad, Takeda, and
  Milanfar}]{protter2008generalizing}
Protter M, Elad M, Takeda H, Milanfar P (2008) Generalizing the nonlocal-means
  to super-resolution reconstruction. IEEE Transactions on image processing
  18(1):36--51

\bibitem[{Rahaman et~al.(2019)Rahaman, Baratin, Arpit, Draxler, Lin, Hamprecht,
  Bengio, and Courville}]{rahaman2019spectral}
Rahaman N, Baratin A, Arpit D, Draxler F, Lin M, Hamprecht F, Bengio Y,
  Courville A (2019) On the spectral bias of neural networks. In: ICML

\bibitem[{Rasti et~al.(2021)Rasti, Koirala, Scheunders, and
  Ghamisi}]{rasti2021undip}
Rasti B, Koirala B, Scheunders P, Ghamisi P (2021) Undip: Hyperspectral
  unmixing using deep image prior. IEEE Transactions on Geoscience and Remote
  Sensing

\bibitem[{Ribes and Schmitt(2008)}]{ribes2008linear}
Ribes A, Schmitt F (2008) Linear inverse problems in imaging. IEEE Signal
  Processing Magazine 25(4):84--99

\bibitem[{Roth and Black(2009)}]{roth2009fields}
Roth S, Black MJ (2009) Fields of experts. International Journal of Computer
  Vision 82(2):205

\bibitem[{Ruderman(1994)}]{ruderman1994statistics}
Ruderman DL (1994) The statistics of natural images. Network: Computation in
  Neural Systems 5(4):517--548

\bibitem[{Rudin et~al.(1992)Rudin, Osher, and Fatemi}]{rudin1992nonlinear}
Rudin LI, Osher S, Fatemi E (1992) Nonlinear total variation based noise
  removal algorithms. Physica D: nonlinear phenomena 60(1-4):259--268

\bibitem[{Sheikh et~al.(2006)Sheikh, Sabir, and Bovik}]{sheikh2006statistical}
Sheikh HR, Sabir MF, Bovik AC (2006) A statistical evaluation of recent full
  reference image quality assessment algorithms. IEEE Transactions on image
  processing 15(11):3440--3451

\bibitem[{Shi et~al.(2021)Shi, Chen, Gavves, Mettes, and
  Snoek}]{shi2021unsharp}
Shi Z, Chen Y, Gavves E, Mettes P, Snoek CG (2021) Unsharp mask guided
  filtering. IEEE Transactions on Image Processing 30:7472 -- 7485

\bibitem[{Simoncelli and Olshausen(2001)}]{simoncelli2001natural}
Simoncelli EP, Olshausen BA (2001) Natural image statistics and neural
  representation. Annual Review of Neuroscience 24(1):1193--1216

\bibitem[{Titterington(1985)}]{titterington1985general}
Titterington D (1985) General structure of regularization procedures in image
  reconstruction. Astronomy and Astrophysics 144:381

\bibitem[{Ulyanov et~al.(2018)Ulyanov, Vedaldi, and
  Lempitsky}]{ulyanov2018deep}
Ulyanov D, Vedaldi A, Lempitsky V (2018) Deep image prior. In: CVPR

\bibitem[{Ulyanov et~al.(2020)Ulyanov, Vedaldi, and
  Lempitsky}]{ulyanov2020deep}
Ulyanov D, Vedaldi A, Lempitsky V (2020) Deep image prior. International
  Journal of Computer Vision 128(7)

\bibitem[{Vu et~al.(2021)Vu, DiSpirito, Li, Wang, Zhu, Chen, Jiang, Zhang, Luo,
  Zhang, Zhou, Horstmeyer, and Yao}]{vu2021deep}
Vu T, DiSpirito A, Li D, Wang Z, Zhu X, Chen M, Jiang L, Zhang D, Luo J, Zhang
  YS, Zhou Q, Horstmeyer R, Yao J (2021) Deep image prior for undersampling
  high-speed photoacoustic microscopy. Photoacoustics 22:100266

\bibitem[{Wan et~al.(2020)Wan, Zhang, Chen, Zhang, Chen, Liao, and
  Wen}]{wan2020bringing}
Wan Z, Zhang B, Chen D, Zhang P, Chen D, Liao J, Wen F (2020) Bringing old
  photos back to life. In: CVPR, pp 2747--2757

\bibitem[{Xu et~al.(2020)Xu, Zhang, Luo, Xiao, and Ma}]{xu2019frequency}
Xu ZQJ, Zhang Y, Luo T, Xiao Y, Ma Z (2020) Frequency principle: Fourier
  analysis sheds light on deep neural networks. Communications in Computational
  Physics

\bibitem[{Zeyde et~al.(2010)Zeyde, Elad, and Protter}]{zeyde2010single}
Zeyde R, Elad M, Protter M (2010) On single image scale-up using
  sparse-representations. In: ICCS

\bibitem[{Zhang et~al.(2017)Zhang, Zuo, Chen, Meng, and
  Zhang}]{zhang2017beyond}
Zhang K, Zuo W, Chen Y, Meng D, Zhang L (2017) Beyond a gaussian denoiser:
  Residual learning of deep cnn for image denoising. IEEE Transactions on Image
  Processing 26(7):3142--3155

\bibitem[{Zhang et~al.(2018)Zhang, Zuo, and Zhang}]{zhang2018ffdnet}
Zhang K, Zuo W, Zhang L (2018) Ffdnet: Toward a fast and flexible solution for
  cnn-based image denoising. IEEE Transactions on Image Processing
  27(9):4608--4622

\bibitem[{Zukerman et~al.(2020)Zukerman, Tirer, and Giryes}]{zukerman2020bp}
Zukerman J, Tirer T, Giryes R (2020) Bp-dip: A backprojection based deep image
  prior. In: EUSIPCO

\end{thebibliography}
